\documentclass[11pt, a4paper]{article}
\usepackage{jcappub}
\usepackage[T1]{fontenc} 

\usepackage{bbm}
\usepackage{slashed}
\usepackage{color}

\usepackage{epsfig,graphicx,bm,color} 
\usepackage{dcolumn}
\usepackage{bm}
\usepackage{hyperref}
\usepackage{color}
\usepackage{breakurl}
\usepackage{soul} 

\usepackage{aas_macros}
\usepackage{tablefootnote}
\usepackage[displaymath]{lineno}
\usepackage{subfigure}
\usepackage{multirow}
\usepackage{amsmath}

\def\rpv{{R}_{p} \hspace{-0.4cm}\slash\hspace{0.2cm}}

\def\lsim{\raise0.3ex\hbox{$\;<$\kern-0.75em\raise-1.1ex\hbox{$\sim\;$}}}
\def\gsim{\raise0.3ex\hbox{$\;>$\kern-0.75em\raise-1.1ex\hbox{$\sim\;$}}}

\newcommand{\captions}{\sf\caption}
\def    \beq            {\begin{equation}}
\def    \eeq            {\end{equation}}
\def    \bea           {\begin{eqnarray}}
\def    \eea           {\end{eqnarray}}

\def \mn{\mu\nu{\rm SSM}}
\def\n{\widetilde\chi^0}

\def\g2{{\rm GeV}^2}

\def\n{\widetilde\chi^0}

\def\sw2{sin^2 \theta_w}

\def\a^tau{\alpha_{\tau}}

\def\beq{\begin{equation}}
\def\eeq{\end{equation}}
\def\beqa{\begin{eqnarray}}
\def\eeqa{\end{eqnarray}}

\newcommand{\tev}{\,\textrm{TeV}}
\newcommand{\gev}{\,\textrm{GeV}}

\newcommand{\newc}{\newcommand}
\newc\BR{BR}
\newc{\akappa}{A_{\kappa} }
\newc\deltagmtwo{\delta (g-2)_{\mu}} 
\newc\deltaamu{\Delta a_{\mu}}

\def\anti{\overline}

\def\rpv{{R}_{p} \hspace{-0.4cm}\slash\hspace{0.2cm}}

\newc{\haa}{BR\(h_1\to a_1 a_1\)}
\newc{\abb}{BR\(a_1\to b\anti{b}\)}
\newc{\hbb}{BR\(h_1\to b\anti{b}\)}
\newc{\Fermi}{\textit{Fermi}-}
\newc{\abund}{\Omega h^2}
\newc\bsgamma{b\rightarrow s \gamma }
\newc\bxsgamma{\overline{B}\rightarrow X_{s}\gamma}
\newc\brbsgamma{\BR(\overline{B}\rightarrow X_s\gamma)}

\begin{document}

\thispagestyle{empty}
\begin{flushright}
FTUAM-16-31; 
IFT-UAM/CSIC-16-079\\
\end{flushright}


\title{Search for sharp and smooth spectral
signatures of $\mu\nu$SSM gravitino dark matter with Fermi-LAT}

\author[a,b]{Germ\'an~A.~G\'omez-Vargas,}
\author[c]{Daniel~E.~L\'opez-Fogliani,}
\author[d,e]{Carlos~Mu\~noz,}
\author[c]{Andres~D.~Perez,}
\author[f]{Roberto~Ruiz~de~Austri}


\affiliation[a]{Instituto de Astrof\'isica, Pontificia Universidad Cat\'olica de Chile, Avenida Vicu\~na Mackenna 4860, Santiago, Chile}
\affiliation[b]{Istituto Nazionale di Fisica Nucleare, Sezione di Roma ``Tor Vergata'', I-00133 Roma, Italy}
\affiliation[c]{Instituto de F\'isica de Buenos Aires, UBA \& CONICET, Departamento de F\'isica, Facultad de Ciencia Exactas y Naturales, Universidad de Buenos Aires, 1428 Buenos Aires, Argentina}
\affiliation[d]{Departamento de F\'isica Te\'orica, Universidad Aut\'onoma de Madrid (UAM), Cantoblanco, 28049 Madrid, Spain}
\affiliation[e]{Instituto de F\'isica Te\'orica UAM-CSIC, Campus de Cantoblanco UAM, 28049 Madrid, Spain}
\affiliation[f]{Instituto de F\'{\i}sica Corpuscular CSIC--UV, c/ Catedr\'atico  Jos\'e Beltr\'an 2, 46980 Paterna (Valencia), Spain}

\emailAdd{ggomezv@uc.cl}
\emailAdd{daniel.lopez@df.uba.ar}
\emailAdd{c.munoz@uam.es}
\emailAdd{andres.perez@df.uba.ar}
\emailAdd{rruiz@ific.uv.es}

\abstract{
The $\mu\nu$SSM solves the $\mu$ problem of supersymmetric models and reproduces neutrino data, simply using couplings with right-handed neutrinos $\nu$'s.
Given that these couplings break explicitly $R$ parity, the gravitino is a natural candidate for decaying dark matter in the $\mn$.
In this work we carry out a complete analysis of the detection of $\mn$ gravitino dark matter through 
$\gamma$-ray observations.
In addition to the two-body decay producing a
sharp line,
we include in the analysis the three-body decays producing a smooth spectral signature.
We perform first a deep exploration of the low-energy parameter space of the $\mn$ taking into account that neutrino data must be reproduced. 
Then, we compare the $\gamma$-ray fluxes predicted by the model with \Fermi LAT observations.
In particular, with the
95$\%$ CL upper limits on the total diffuse extragalactic $\gamma$-ray background using 50 months of data, together with the upper limits on line emission from an updated analysis using
69.9 months of data.
For standard values of bino and wino masses,
gravitinos with masses larger than about 4 GeV, or lifetimes smaller
than about $10^{28}$ s, produce too large fluxes and are excluded as dark matter candidates. 
However, when limiting scenarios with large and close values of the gaugino masses are considered,
the constraints turn out to be less stringent, excluding masses larger than 17 GeV and lifetimes smaller
than $4\times 10^{25}$ s.
}

\maketitle

\section{Introduction}\label{sec:intro}

The 
`$\mu$ from $\nu$' supersymmetric standard model 
($\mu\nu$SSM) \cite{propuvSSM,analisisparam},
introduces couplings with right-handed (RH) neutrino superfields in the superpontential
in order to solve the 
$\mu$-problem, while simultaneously explain the origin of neutrino masses.
In particular,
the 
couplings
$\lambda_i \hat \nu^c_i \hat H_d\hat H_u$ generate an
effective $\mu$ term through RH sneutrino vacuum expectation values (VEVs),
$\langle \tilde \nu^c_i \rangle \equiv v_{\nu^c_i}$, after the successful 
electroweak symmetry breaking (EWSB): $\mu^{eff}=\lambda_i v_{\nu^c_i}$.
In addition, the couplings $\frac{1}{3}\kappa{_{ijk}} \hat \nu^c_i\hat \nu^c_j\hat \nu^c_k$
generate effective Majorana masses for the RH neutrinos,
$M_{ij}^{eff}=2\kappa_{ijk}v_{\nu^c_k}$, giving rise to a generalized electroweak-scale seesaw mechanism
which can reproduce the observed neutrino masses and mixing angles.
On the other hand, these couplings violate $R$ parity explicitly ($\rpv$), 
and therefore the lightest supersymmetric particle (LSP) is not stable,
implying that the phenomenology of the 
$\mu\nu$SSM\footnote{For reviews about the LHC phenomenology of the $\mn$, 
see~\cite{Munoz:2009an,LopezFogliani:2010bf,reviewsmunu3} and references therein.} is very different from the one of the 
minimal supersymmetric standard model (MSSM) or the one of the the next-to-MSSM (NMSSM).

Given the interest of the $\mn$ as an alternative to the usual supersymmetric models,
solving crucial problems and generating a different phenomenology, 
it is worth studying candidates for dark matter (DM) in this context and their possible signals.

The usual DM candidates in the case of $R$-parity conserving models such as the MSSM or the NMSSM, i.e. the neutralino~\cite{Goldberg:1983nd,Ellis:1983wd,Krauss:1983ik,Ellis:1983ew} or the RH sneutrino (see e.g. Ref.~\cite{Cerdeno:2008ep} and references therein), have very short lifetimes in $\rpv$ models and therefore can no longer be used.
Nevertheless, if the
gravitino is the LSP, it
can be a good candidate for DM,
since its lifetime turns out to be much longer than the age of the Universe, being
suppressed both by
the gravitational interaction and by the typically small $\rpv$ couplings~\cite{Takayama:2000uz}.
In addition, since it decays producing $\gamma$ rays, gravitino DM could be detected in $\gamma$-ray experiments, 
as discussed in 
Refs.~\cite{Takayama:2000uz,Buchmuller:2007ui,Bertone:2007aw,Ibarra:2007wg,Ishiwata:2008cu,Choi:2010xn,Choi:2010jt,Diaz:2011pc,Restrepo:2011rj} 
in the context of bilinear or trilinear $\rpv$.

In the context of the $\mn$, 
the search for indirect signals of gravitino DM in the $\gamma$-ray sky has been focused on looking for sharp spectral features as it typically decays into a two-body final state, 
photon neutrino ($\gamma\nu$), that gives rise to a mono-energetic $\gamma$ signal~\cite{Choi:2009ng,GomezVargas:2011ph,Albert:2014hwa}.
The non-observation of such a sharp spectral signature with $\gamma$-ray detectors set stringent limits on the gravitino lifetime and mass with important implications in the $\mu\nu$SSM framework.
In particular, assuming that the whole Galactic DM halo is made of $\mu\nu$SSM gravitino particles,
the limits on line emission from \Fermi LAT imply that the gravitino mass has to be lower than about 5 GeV with lifetime larger than about $10^{28}$ s~\cite{Albert:2014hwa}. 

However, gravitino decays through off-shell particles into three-body final states, produces a smooth spectrum of energetic $\gamma$-rays that could also be observed at \Fermi LAT, as pointed out in Refs.~\cite{Choi:2010xn,Choi:2010jt,Diaz:2011pc,Cottin:2014cca}. 
In this paper we perform a deep exploration of the $\mu\nu$SSM parameter space updating previous constraints,
and paying special attention to regions
where the gravitino decays with a sizable branching ratio (BR) into three-body final states, while suppressing the monochromatic photon signal. We probe these $\mu\nu$SSM regions against updated limits for spectral lines~\cite{Ackermann:2015lka} and the latest determination of the extragalactic $\gamma$-ray background (EGB)~\cite{Ackermann:2014usa}, both from \Fermi LAT data.
We use the EGB for analysing smooth gravitino signals because the Region of Interest (ROI) used for extracting the EGB is similar to the optimized ROI for decaying dark matter searches analyzed in \cite{Ackermann:2013uma} (see appendix B of that reference for details). 
We will show 
that previous constraints can be relaxed, with an upper bound for the gravitino mass of 17 GeV and a lower bound on the lifetime of $4\times 10^{25}$~s.

This work is organized as follows. 
In Section~\ref{model}, we introduce the $\mn$ and explain how neutrino data can easily be reproduced in the model.
In Section~\ref{sec:Scan}, we discuss the two- and three-body gravitino decay channels, and present the $\mu\nu$SSM low-energy parameter space region that suppresses the gravitino decay into $\gamma\nu$ with large BRs into three-body channels. In Section~\ref{sec:gamma}, signals from  decaying gravitinos are first analyzed.
Then, we present the $\gamma$-ray measurements by \Fermi LAT employed to probe 
the $\mu\nu$SSM parameter space region discussed in Section~\ref{sec:Scan}. 
Finally, we present our results in Section~\ref{sec:results}. The conclusions are left for 
Section~\ref{sec:conclusions}.

\section{The $\mu\nu$SSM and neutrino physics}
\label{model}

The superpotential of the $\mu\nu$SSM 
contains in addition to the MSSM
Yukawas for quarks and charged leptons,
Yukawas for neutrinos, and the two couplings discussed in the introduction that
generate the effective $\mu$ term and Majorana masses~\cite{propuvSSM,analisisparam}, producing as well explicit $\rpv$:
\begin{eqnarray}
W & = &
\ \epsilon_{ab} \left(
Y_{u_{ij}} \, \hat H_u^b\, \hat Q^a_i \, \hat u_j^c +
Y_{d_{ij}} \, \hat H_d^a\, \hat Q^b_i \, \hat d_j^c +
Y_{e_{ij}} \, \hat H_d^a\, \hat L^b_i \, \hat e_j^c +
Y_{\nu_{ij}} \, \hat H_u^b\, \hat L^a_i \, \hat \nu^c_j 
\right)
\nonumber\\
& - &
\epsilon{_{ab}} \lambda_{i} \, \hat \nu^c_i\,\hat H_d^a \hat H_u^b
+
\frac{1}{3}
\kappa{_{ijk}} 
\hat \nu^c_i\hat \nu^c_j\hat \nu^c_k \ .
\label{superpotential}
\end{eqnarray}
Since only dimensionless trilinear couplings are present in 
(\ref{superpotential}), the EWSB
is determined by the usual soft supersymmetry-breaking terms of the scalar potential. 
In addition to the soft terms, the neutral scalar potential
receives the $D$ and $F$ term contributions 
that can be found in Refs.~\cite{propuvSSM,analisisparam}.
With the choice of CP conservation,\footnote{$\mu\nu$SSM with spontaneous 
CP violation was studied in Ref.~\cite{neutrinocp}.} after the EWSB the neutral scalars
develop in general the following real VEVs:
\begin{equation}
\langle H_d^0 \rangle = v_d, \, \quad \langle H_u^0 \rangle = v_u, \,
\quad \langle \widetilde \nu_i \rangle = v_{\nu_i}, \,  \quad
\langle \widetilde \nu_i^c \rangle = v_{\nu^c_i}\ ,
\label{vevs}
\end{equation}
%
where in addition to the usual ones of the MSSM Higgses $H_u^0$ and $H_d^0$, the new couplings generate VEVs for left-handed (LH) sneutrinos $\widetilde \nu_i$, as well as for the right-handed (RH) sneutrinos $\widetilde \nu_i^c$.


The VEVs of the RH sneutrinos, $v_{\nu_{j}^{c}}$, are naturally of the order of the 
EWSB scale~\cite{propuvSSM}, confirming that the $6^{\rm th}$ term in the superpotential (\ref{superpotential})
generates the effective Majorana masses for RH neutrinos, as discussed in the Introduction.
Thus we can implement naturally an electroweak-scale seesaw
in the $\mu\nu$SSM, asking for 
neutrino Yukawa 
couplings of the order of the electron Yukawa coupling or smaller,
$Y_{\nu_{ij}} \sim 10^{-6} - 10^{-7}$
~\bibpunct{}{}{,}{n}{}{}[\cite{propuvSSM}, \cite{analisisparam}, \cite{Ghosh:2008yh}, \cite{Bartl:2009an}, \cite{neutrinocp}, \cite{Ghosh:2010zi}, \cite{LopezFogliani:2010bf}, \cite{Ghosh:2010ig}]\bibpunct{[}{]}{,}{n}{}{},
i.e. we work with Dirac masses for neutrinos, $m_D\sim Y_{\nu}v_u\lsim 10^{-4}$ GeV.
On the other hand, the VEVs 
of the LH sneutrinos,
$v_{\nu_i}$, are much smaller than the other VEVs (\ref{vevs}) in the $\mu\nu$SSM. This is 
because of their minimization conditions, where the contributions of $Y_{\nu_{ij}}$ are relevant implying
$v_{\nu}\to 0$ as $Y_{\nu}\to 0$.
It is then easy to estimate the values of $v_{\nu}$ as $v_{\nu}\lsim m_D$~\cite{propuvSSM}.

For our computation below we are interested in the neutral fermion mass matrix of the $\mn$.
In this model there are new couplings and VEVs (see Eqs.~(\ref{superpotential}) and (\ref{vevs})), implying larger mass matrices than those of the MSSM/NMSSM.
In particular, in the case of the neutralinos, they turn out to be also mixed 
with the LH and RH neutrinos. Besides, we saw before that Majorana masses for RH neutrinos are generated dynamically, thus they will behave as the singlino components of the neutralinos.
Altogether, in a basis where
${\chi^{0}}^T=(\tilde{B^{0}},
\tilde{W^{0}},\tilde{H^0_{d}},\tilde{H^0_{u}},\nu_{R_i},\nu_{L_i})$,
one obtains the following $10\times 10$ neutral fermion
(neutralino-neutrino) mass
matrix~\cite{propuvSSM,analisisparam}:
%
\begin{equation}
{\mathcal M}_n=
\left(
\begin{array}{cc}
M & m\\
m^{T} & 0_{3\times3}
\end{array}
\right)\ ,
\label{mixing}
\end{equation}
with

\begin{equation}
M=
\left(
\begin{array}{ccccccc}
M_{1} & 0 & -A v_{d} & A v_{u} & 0 & 0 & 0\\
0 & M_{2} &B v_{d} & -B v_{u} & 0 & 0 & 0\\
-A v_{d} & B v_{d} & 0 & -\lambda_{i}v_{\nu^c_{i}} & -\lambda_{1}v_{u} & -\lambda_{2}v_{u} & -\lambda_{3}v_{u}\\
A v_{u} & -B v_{u} & \: \: -\lambda_{i}v_{\nu^c_{i}} & 0 & -\lambda_{1}v_{d}+Y_{\nu_{i1}}v_{\nu_{i}} & -\lambda_{2}v_{d}+Y_{\nu_{i2}}v_{\nu_{i}} & -\lambda_{3}v_{d}+Y_{\nu_{i3}}v_{\nu_{i}}\\
0 & 0 &  -\lambda_{1}v_{u} & \: \:-\lambda_{1}v_{d}+Y_{\nu_{i1}}v_{\nu_{i}} & 2\kappa_{11j}v_{\nu^c_{j}} & 2\kappa_{12j}v_{\nu^c_{j}} & 2\kappa_{13j}v_{\nu^c_{j}}\\
0 & 0 & -\lambda_{2}v_{u} &  \: \: -\lambda_{2}v_{d}+Y_{\nu_{i2}}v_{\nu_{i}} & 2\kappa_{21j}v_{\nu^c_{j}} & 2\kappa_{22j}v_{\nu^c_{j}} & 2\kappa_{23j}v_{\nu^c_{j}}\\
0 & 0 & -\lambda_{3}v_{u} & \: \:-\lambda_{3}v_{d}+Y_{\nu_{i3}}v_{\nu_{i}} & 2\kappa_{31j}v_{\nu^c_{j}} & 2\kappa_{32j}v_{\nu^c_{j}} & 2\kappa_{33j}v_{\nu^c_{j}}
\end{array}
\right)\ ,
\label{neumatrix}
\end{equation}
where $A\equiv\frac{G}{\sqrt{2}} \sin\theta_W$, $B\equiv\frac{G}{\sqrt{2}} \cos\theta_W$,
with $G^2\equiv g_{1}^{2}+g_{2}^{2}$,
and
\begin{equation}
m^{T}=
\left(\begin{array}{ccccccc}
-\frac{g_{1}}{\sqrt{2}}v_{\nu_{1}} \: & \: \frac{g_{2}}{\sqrt{2}}v_{\nu_{1}} & \: 0 & \: Y_{\nu_{1i}}v_{\nu^c_{i}} & \: Y_{\nu_{11}}v_{u} & \: Y_{\nu_{12}}v_{u} & \: Y_{\nu_{13}}v_{u}\\
\: -\frac{g_{1}}{\sqrt{2}}v_{\nu_{2}} & \: \frac{g_{2}}{\sqrt{2}}v_{\nu_{2}} & \: 0 & \: Y_{\nu_{2i}}v_{\nu^c_{i}} & \: Y_{\nu_{21}}v_{u} & \: Y_{\nu_{22}}v_{u} & \: Y_{\nu_{23}}v_{u}\\
\: -\frac{g_{1}}{\sqrt{2}}v_{\nu_{3}}\: & \: \frac{g_{2}}{\sqrt{2}}v_{\nu_{3}} & \: 0 & \: Y_{\nu_{3i}}v_{\nu^c_{i}} & \: Y_{\nu_{31}}v_{u} & \: Y_{\nu_{32}}v_{u} & \: Y_{\nu_{33}}v_{u}\end{array}\right)\ .
\label{mixing3}
\end{equation}

\noindent The structure of this mass matrix is that of a 
generalized electroweak-scale seesaw, since it involves not only the RH neutrinos but also the neutralinos. 
Because of this structure, data on neutrino physics can easily be 
reproduced at tree level~\bibpunct{}{}{,}{n}{}{}[\cite{propuvSSM}, \cite{analisisparam}, \cite{Ghosh:2008yh}, \cite{neutrinocp}, \cite{Ghosh:2010zi}]\bibpunct{[}{]}{,}{n}{}{},
even with diagonal Yukawa couplings \cite{Ghosh:2008yh,neutrinocp}, i.e.
$Y_{\nu_{ii}}=Y_{\nu_i}$ and vanishing otherwise.
Qualitatively, we can understand all this in the following way. First of all, neutrino masses are going to be very small since
the entries of the matrix $M$ are much larger than the ones of the matrix $m$. Notice in this respect that the entries of $M$ are of the order of the electroweak scale, whereas the ones in $m$ are of the order 
of the Dirac masses for neutrinos~\cite{propuvSSM,analisisparam}. 
Second, from the above matrices, in the limit of large $\tan\beta$ 
(where $\tan\beta\equiv v_u/v_d$) one can obtain a simplified formula for the effective neutrino mixing mass 
matrix~\cite{neutrinocp}:
\begin{eqnarray}
(m^{eff}_{\nu})_{ij} 
\simeq \frac{Y_{\nu_i}Y_{\nu_j}v_u^2}
{6 \kappa v_{\nu^c}}
                   (1-3 \delta_{ij})-\frac{v_{\nu_i} v_{\nu_j}}{2M}  \ ,
\label{Limit no mixing Higgsinos gauginos}
\end{eqnarray}
where $\kappa_{iii}\equiv \kappa_{i} \equiv \kappa$ and vanishing otherwise,
$v_{\nu^c_i}\equiv v_{\nu^c}$, and
${M} \equiv \frac{M_1 M_2}{g_1^2 M_2 + g_2^2 M_1}$. 
Using this formula it is easy
to understand how diagonal Yukawas $Y_{\nu_i}$ can give rise to off-diagonal entries in the mass matrix. One of the key points is the extra contribution given by the first term of 
Eq.~(\ref{Limit no mixing Higgsinos gauginos}) with respect to the ordinary seesaw where it is absent. Another extra contribution to the off-diagonal entries is the third term generated through the mixing of LH neutrinos with gauginos.

\section{Gravitino decay}
\label{sec:Scan}

The gravitino LSP is an interesting candidate for DM in
$\rpv$ models. 
The gravitino has an interaction term in the supergravity Lagrangian
with 
the photon and the photino. As discussed in the previous section, 
in the presence of $\rpv$ couplings
the photino and the LH neutrinos can be mixed in the neutral fermion mass matrix~(see Eq.~(\ref{mixing3}) for the case of the $\mn$), and therefore
the gravitino will be able to decay 
through the interaction term
into a photon and a neutrino \cite{Takayama:2000uz}, as shown in Fig.~\ref{fig_gfn}. 
This has important implications because
the $\gamma$-ray signal is a sharp line with an energy
$\frac{m_{3/2}}{2}$, that can be detected in $\gamma$-ray satellite experiments, such as \Fermi LAT.
The result for the decay width is given by \cite{Takayama:2000uz}:
\begin{figure}[t]
 \begin{center}
       \epsfig{file=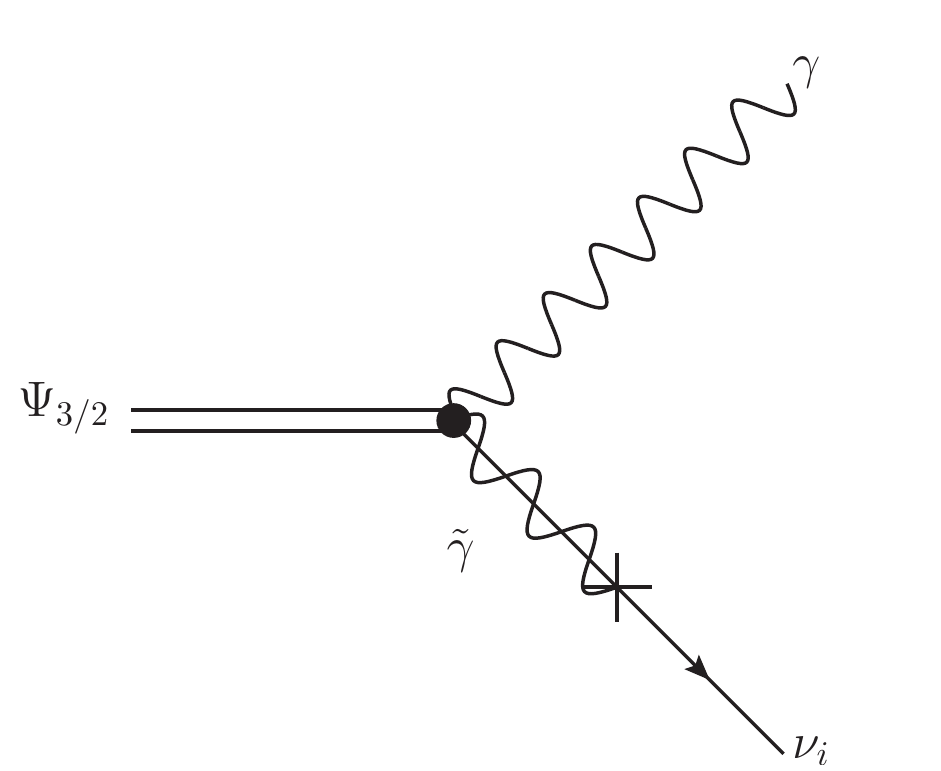,height=4.1cm}
\captions{Tree-level diagram for the two-body decay of a gravitino into a photon and a neutrino, via 
photino-neutrino mixing.}
    \label{fig_gfn}
 \end{center}
\end{figure}
\bea
\Gamma(\Psi_{3/2}\rightarrow\sum_i\gamma\nu_i)\simeq\frac{m_{3/2}^3}{64\pi M_{P}^2}|U_{\tilde{\gamma} \nu}|^2\ ,
\label{decay2body}
\eea
where $m_{3/2}$ is the gravitino mass, $M_{P}\simeq 2.4\times 10^{18}\gev$ is the reduced Planck mass, and the mixing parameter $|U_{\tilde{\gamma} \nu_i}|$ determines the photino content of the neutrino 
\bea
\left|U_{\tilde{\gamma} \nu}\right|^2= \sum^3_{i=1}\left|N_{i1} \, \cos\theta_W +  N_{i2} \, \sin\theta_W\right|^2.
\label{photino}
\eea
Here $N_{i1} (N_{i2})$ is the bino (wino) component of the $i$-th neutrino, and $\theta_{W}$ is the weak mixing angle. The same result for the decay width holds for the conjugated 
processes $\Psi_{3/2}\rightarrow\gamma\bar{\nu}_i$.

Assuming that this is the only decay channel of the gravitino, its lifetime can then be written as 
\begin{equation}
{\tau}_{3/2}
(\Psi_{3/2}\rightarrow\sum_i\gamma\nu_i
)
=\frac{1}{2\Gamma\left(\Psi_{3/2}\rightarrow\sum_i\gamma\nu_i\right)}
\simeq 3.8\times 10^{27}\, {s}
\left(\frac{10^{-16}}{|U_{\widetilde{\gamma}\nu}|^2}\right)
\left(\frac{10\, \mathrm{GeV}}{m_{3/2}}\right)^{3}\ ,
\label{lifetimegamma}
\end{equation}
where the factor 2 takes into account the charged conjugated final states.

\begin{figure}[t]
\begin{center}
\begin{tabular}{lll}
\hspace*{-2mm}\epsfig{file=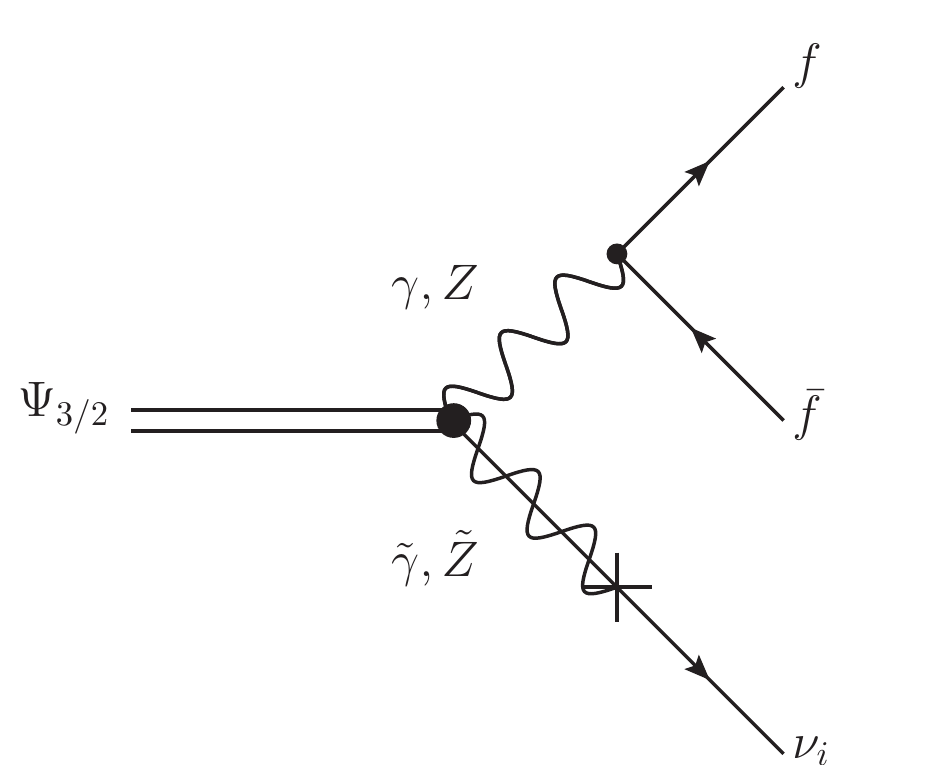,height=4.1cm} & \epsfig{file=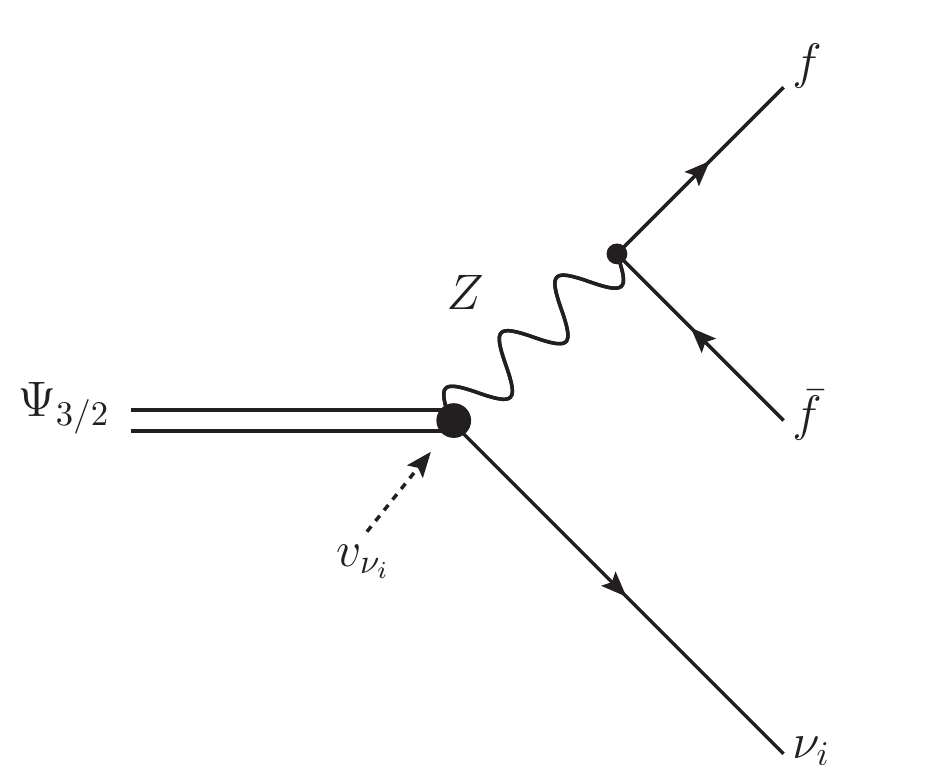,height=4.1cm} & \epsfig{file=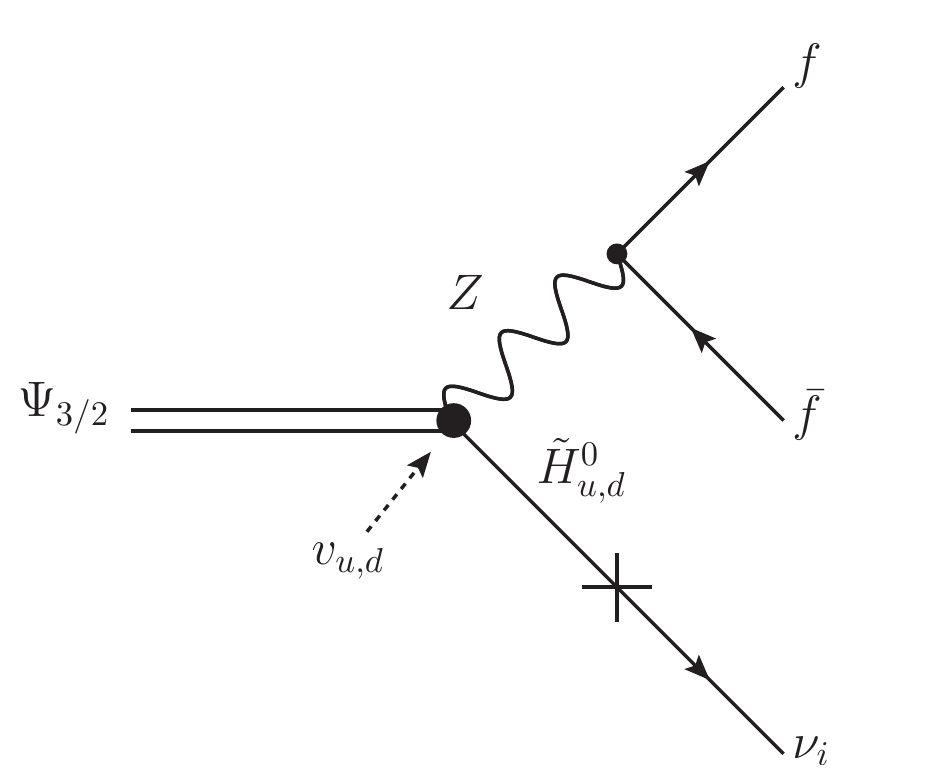,height=4.1cm}\\
\hspace*{2cm}(a) & \hspace*{2cm}(b) & \hspace*{2cm}(c)
\end{tabular}
\end{center}
\captions{Tree-level diagrams for the decay of a gravitino into a fermion-antifermion pair and a neutrino, via an intermediate photon or $Z$ boson.}
    \label{fig_gz}
\end{figure}

\begin{figure}[t]
\begin{center}
\begin{tabular}{lll}
\hspace*{-2mm}\epsfig{file=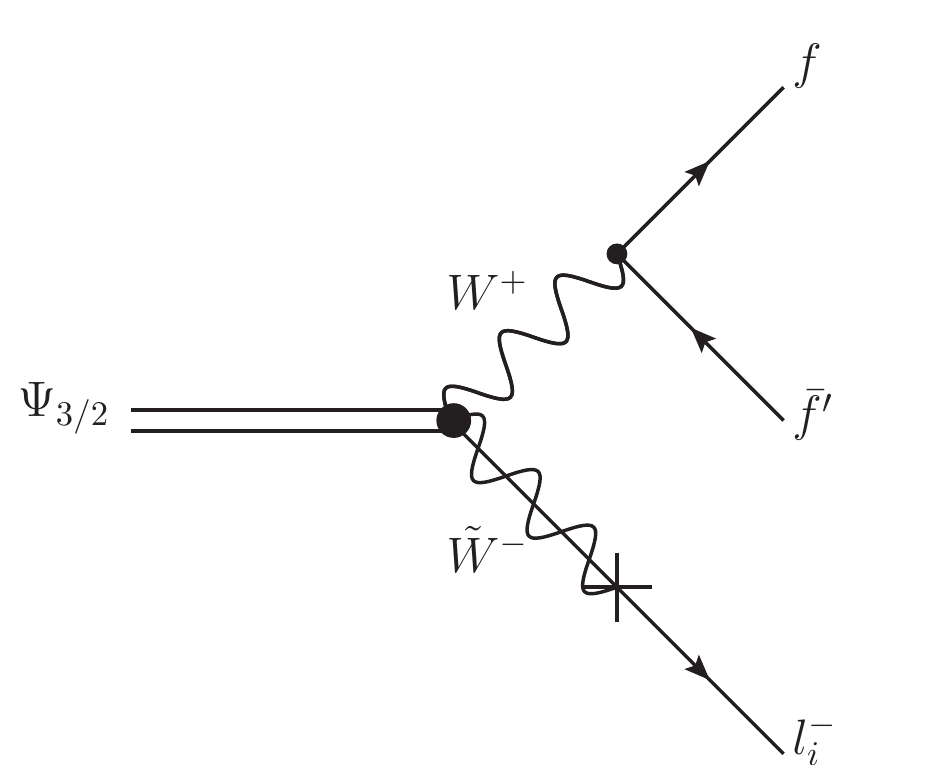,height=4.1cm} & \epsfig{file=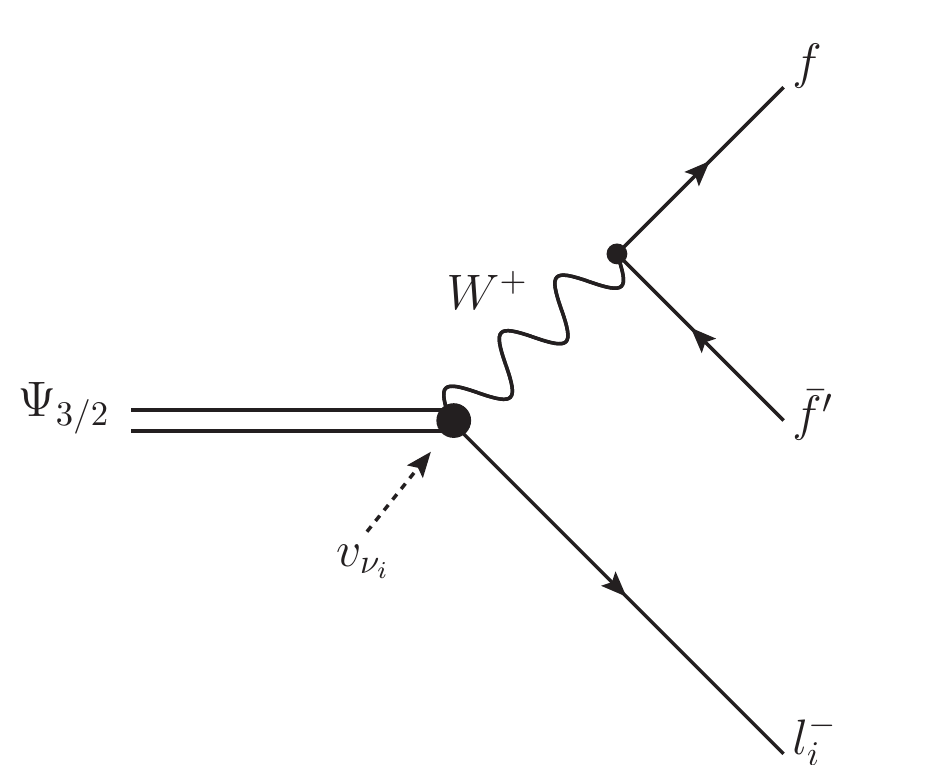,height=4.1cm} & \epsfig{file=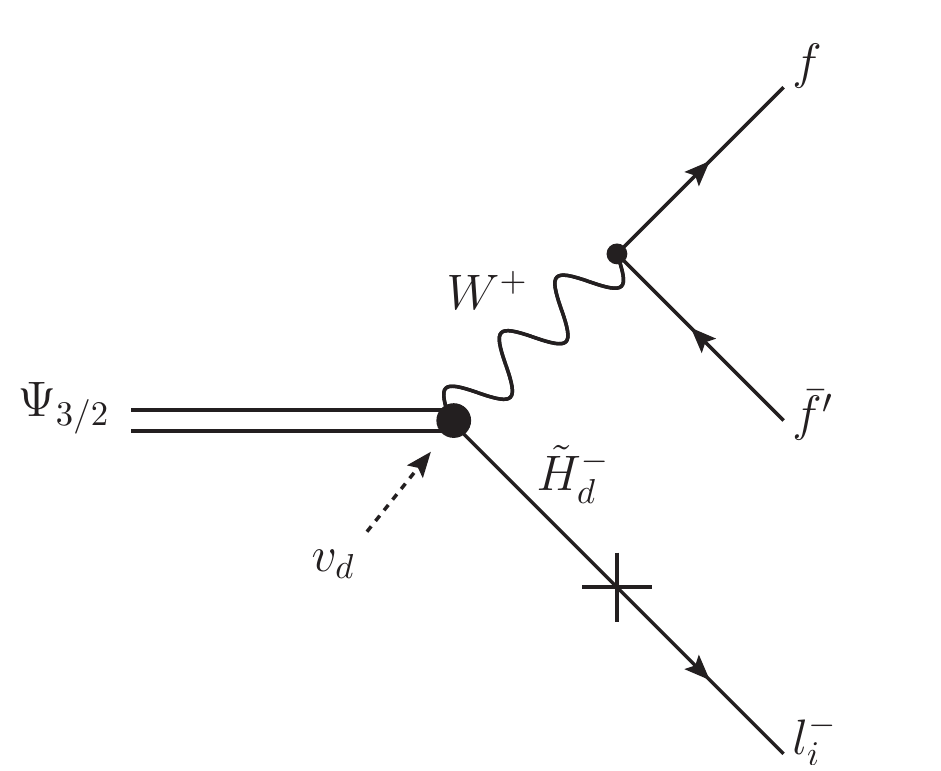,height=4.1cm}\\
\hspace*{2cm}(a) & \hspace*{2cm}(b) & \hspace*{2cm}(c)
\end{tabular}
\end{center}
\captions{Tree-level diagrams for the decay of a gravitino into two fermions and a charged lepton, via an intermediate $W$ boson.}
    \label{fig_gw}
\end{figure}

However, 
as pointed out in Ref.~\cite{Choi:2010xn}, gravitinos with masses smaller than the $W$ mass as in our case
can also decay with a sizable BR into three-body final states, producing
a smooth spectrum of energetic $\gamma$ rays that can also be detected in \Fermi LAT.
These channels are
$\Psi_{3/2}\rightarrow\gamma^*/Z^* \, \nu_i\rightarrow f \,  \bar{f} \, \nu_i$ via an intermediate photon or $Z$ boson, as well as $\Psi_{3/2}\rightarrow W^* \, l\rightarrow f  \, \bar{f}' \, l$ 
via and intermediate $W$ boson, where $f$ denotes fermions and $l$ leptons. 
Both channels are shown in Figs.~\ref{fig_gz} and~\ref{fig_gw}, respectively.
The associated decays were computed in 
Refs.~\cite{Choi:2010xn,Choi:2010jt,Grefe:2011dp,Diaz:2011pc}, and we show for completeness the results of the 
differential decay widths in the Appendix. 
The total gravitino decay width is then given by:
\bea
\Gamma_{\text{total}}(\Psi_{3/2})=2\left(\Gamma(\Psi_{3/2} \rightarrow \gamma\nu) +\Gamma(\Psi_{3/2} \rightarrow f\bar{f}\nu)+\Gamma(\Psi_{3/2} \rightarrow f\bar{f'}l) \right)
\ .
\label{total_width}
\eea
If the $\gamma\nu$ channel is not always the dominant gravitino decay channel, the lifetime
(\ref{lifetimegamma})
gets modified. Obviously, the complete formula becomes 
\begin{equation}
\tau_{3/2}(\Psi_{3/2})
=\frac{1}{\Gamma_{\text{total}}(\Psi_{3/2})}\ .
\label{lifetimetotal}
\end{equation}

As it is easy to understand from the above discussion, 
the photino-neutrino mixing parameter $|U_{\tilde{\gamma} \nu_i}|$ 
plays a crucial role
in the analysis of gravitino DM detection via $\gamma$-ray lines.
We can see from Eq.~(\ref{decay2body}) that, the larger (smaller) $|U_{\tilde{\gamma} \nu_i}|$ the larger (smaller) is the decay width, and therefore more (less) stringent are the constraints on the parameter space of the model from the non-observation of lines in \Fermi LAT data.

On the other hand, in the regions of the parameter space where $|U_{\tilde{\gamma} \nu_i}|$ is suppressed,
the BR to $\gamma\nu$ is also suppressed, and the BRs 
to three-bodies become more important. The three-body decay rates contain terms independent on the mixing parameters (and proportional to $v_{\nu_{i}}$) that can dominate the total rate in some limits, as shown in the Appendix (and Figs.~\ref{fig_gz} and~\ref{fig_gw}).
This can have two effects. First, it can affect significantly the result for the lifetime, as already mentioned above. Second, the constraints on the parameter space from lines are less stringent, but new constraints might appear from the analysis of the smooth spectrum generated by three-body final states, using \Fermi LAT data.

In the next subsection we will study all these crucial issues for the analysis of gravitino DM and its detection.

\subsection{The photino-neutrino mixing parameter in the $\mn$}


We can easily estimate the value of $|U_{\widetilde{\gamma}\nu}|^{2}$
in the $\mu\nu$SSM~\cite{Choi:2009ng}. Using the mass insertion technique,
from the entries in the neutral fermion mass matrix (\ref{mixing3})
and Fig.~\ref{fig_gfn},
we can deduce that the relevant coupling for the mixing between the photino and the neutrinos is given approximately by $g_{1}v_{\nu}$, and as a consequence
\begin{equation}
|U_{\widetilde{\gamma}\nu}|\sim \frac{g_{1}v_{\nu}}{M_{1}}\ .
\label{representa}
\end{equation}
For typical electroweak-scale values for $M_1$, and 
$v_{\nu}\lsim 10^{-4}$ GeV as discussed in the previous section,
one obtains approximately that in order to reproduce the observed neutrino masses and mixing angles, the photino-neutrino mixing parameter is in the range
\begin{equation}
10^{-15} \lesssim |U_{\widetilde{\gamma}\nu}|^{2} \lesssim 10^{-14}.
\label{representative}
\end{equation}
This was confirmed performing a scan of the low-energy parameter space of the model in Ref.~\cite{Choi:2009ng}.
Now, from the non-observation of prominent sharp features in the diffuse emission measurement reported by the \Fermi LAT Collaboration, limits on the parameter space of the model were 
obtained in Refs.~\cite{Choi:2009ng,Albert:2014hwa}, using the bounds (\ref{representative}).
In the most recent work~\cite{Albert:2014hwa}, some of the authors in collaboration with 
\Fermi LAT members obtained the constraints on the
gravitino mass and lifetime, $m_{3/2}\lsim 2.5$ GeV and $\tau_{3/2}\gsim 10^{28}$ s, where the former (latter) arises from the lower (upper) bound in (\ref{representative}).

However, on the basis of Eq.~(\ref{photino}), it was
also suggested in~\cite{Choi:2009ng} to
relax the lower bound by one order of magnitude\footnote{Note that from these values 
of $|U_{\widetilde{\gamma}\nu}|$ and 
Eq.~(\ref{lifetimegamma}), one can deduce that the gravitino can be very long lived 
compared to the current age of the Universe of about $10^{17}$\,s, and therefore a DM candidate.
In this regard, let us also remember that adjusting the reheating temperature one can reproduce the correct relic density
for each possible value of the gravitino mass~\cite{Bolz:2000fu}.}
\begin{equation}
10^{-16} \lesssim |U_{\widetilde{\gamma}\nu}|^{2} \lesssim 10^{-14}\ .
\label{relaxing}
\end{equation}
From (\ref{photino}), one can infer
that values of the mixing parameter close to zero would be achievable through a cancellation between the bino and wino 
contributions, enlarging therefore the allowed values of $m_{3/2}$.
The analysis of~\cite{Albert:2014hwa} under the assumption (\ref{relaxing}) obtained the bound, 
$m_{3/2}\lsim 5$ GeV.


In this work we want to check this assumption in a quantitative way, given its importance when studying the
constraints on the parameter space.
To understand first the situation qualitatively, we will use the result of Refs.~\cite{Ibarra:2007wg,Grefe:2011dp,Diaz:2011pc}, where a more precise approximation for the value of the mixing parameter was obtained. We can recover
that result using again the entries of the neutral fermion mass matrix (\ref{mixing3}), to write
$N_{i1} \approx \frac{-g_1 v_{\nu_i}}{\sqrt 2 M_1} $ and 
$N_{i2} \approx \frac{ g_2 v_{\nu_i}}{\sqrt 2 M_2}$
in Eq.~(\ref{photino}). Thus we obtain:
%
\begin{equation}
	U_{\tilde{\gamma} \nu_i}\approx -\frac{g_1}{\sqrt 2} v_{\nu_i} \cos \theta_W \frac{M_2-M_1}{M_1M_2}\ .
	\label{UAP1}
\end{equation}
It is now trivial to realize that one can suppress the gravitino decay into $\gamma\nu$ canceling the numerator
by simply taking $M_2 \to M_1$.

From Eq.~(\ref{UAP1}), one can also deduce that another way to obtain a small
photino composition of the neutrinos is to increase the denominator using large values of $|M_1|$ and $|M_2|$.
Let us remark nevertheless that the parameters involved in this equation, gauginos masses and VEVs of LH sneutrinos, are also involved in the generalized electroweak-scale seesaw that generates neutrino masses in the $\mn$ (see e.g. the approximate formula (\ref{Limit no mixing Higgsinos gauginos})). Since the values chosen for the parameters must reproduce current data on neutrino masses and mixing angles, we must take into account in the analysis the possible correlations among them.
For example, from the second term in (\ref{Limit no mixing Higgsinos gauginos}), we can see that given a set of parameters that reproduce the neutrino physics, if we increase the values of $|M_{1,2}|$ by two orders of magnitude we have to increase also the LH sneutrino VEVs $v_{\nu_{i}}$ by one order of magnitude. Otherwise, the neutrino physics would be modified.
Thus the photino content of the neutrino, $|U_{\tilde{\gamma} \nu_i}|$, decreases only by one order of magnitude according to Eq.~(\ref{UAP1}) (or Eq.~(\ref{representa})). 
This seems to imply that in order to decrease the value of $|U_{\tilde{\gamma} \nu_i}|$, the strategy of making the gaugino masses similar will be more efficient than increasing their absolute values.

\begin{figure}[t!]
 \begin{tabular}{cc}
 \hspace*{-4mm}
 \epsfig{file=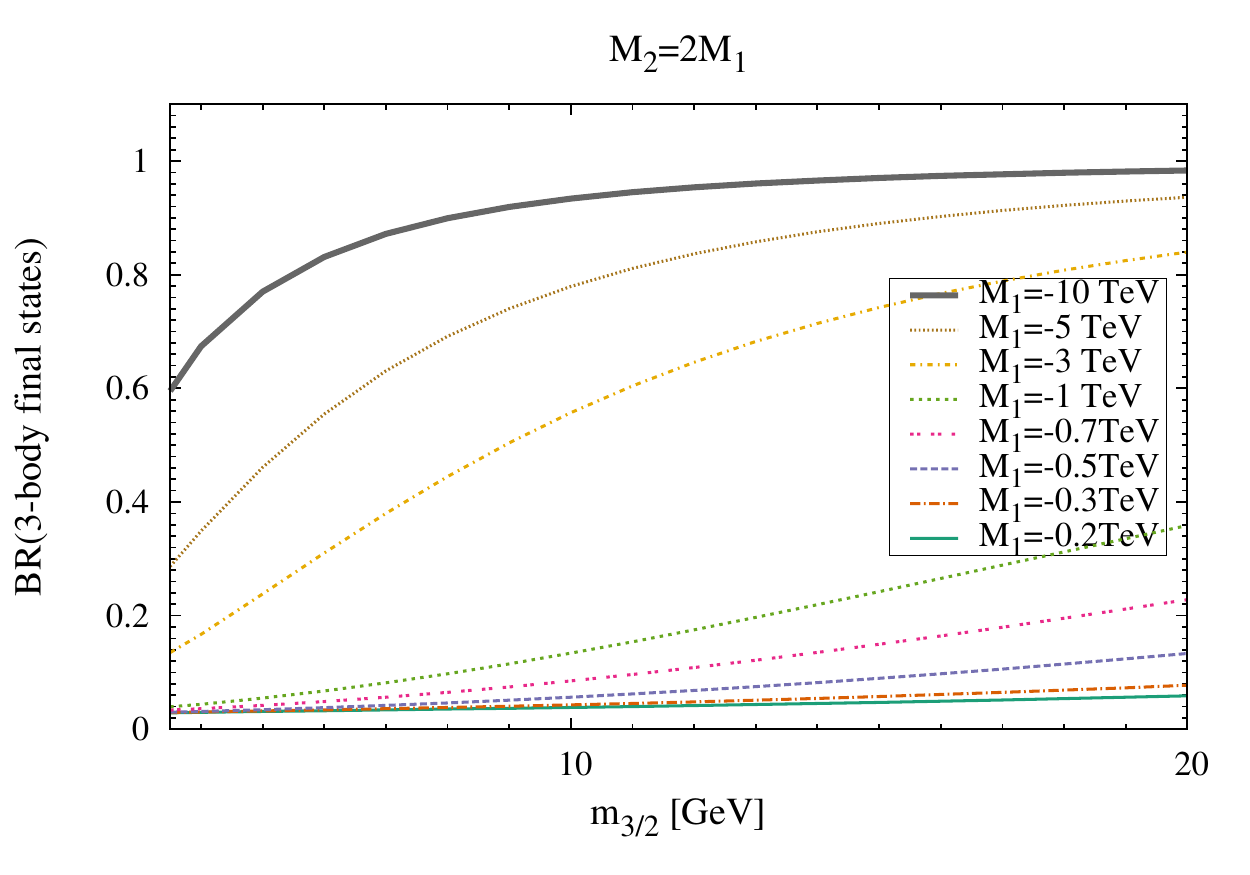,height=5.4cm} 
       \epsfig{file=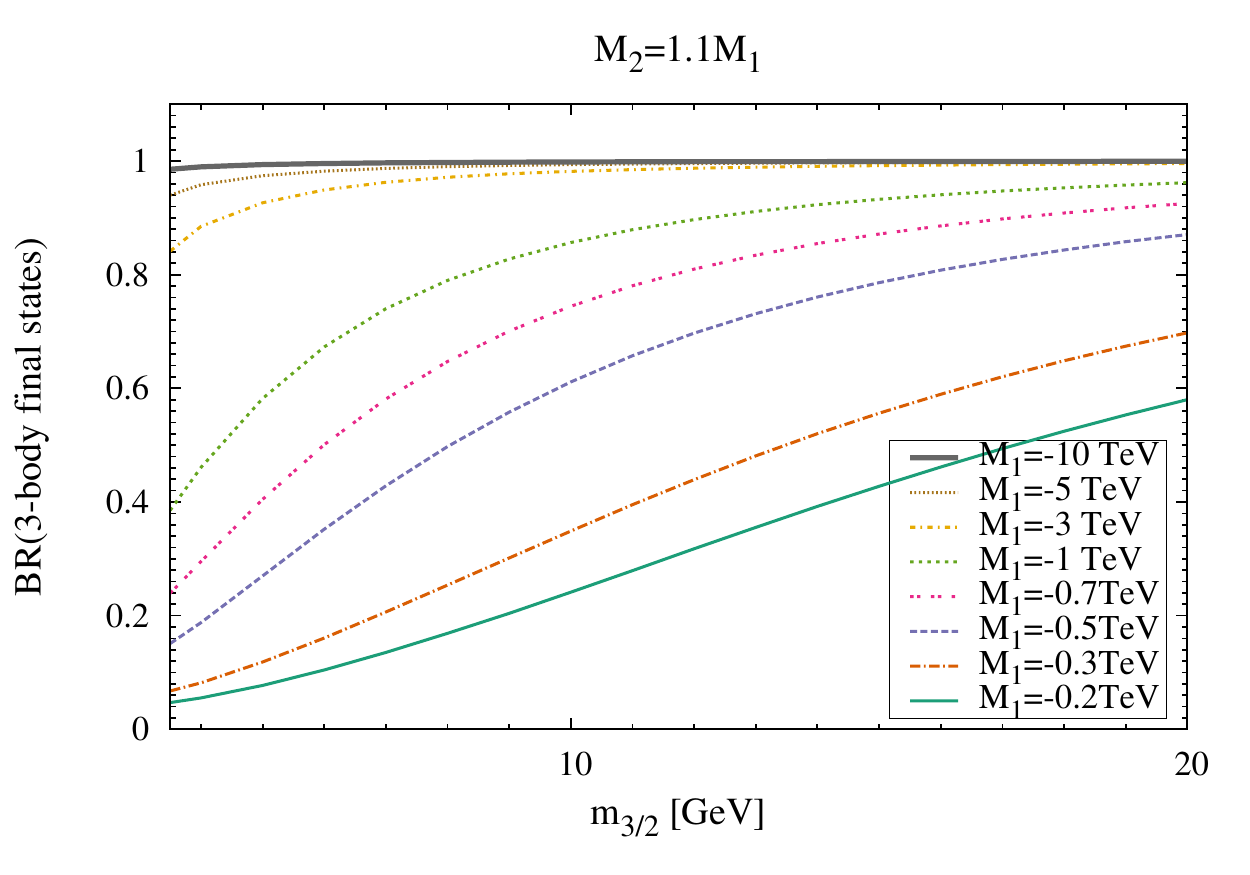,height=5.4cm}   
       \vspace*{-0.8cm}       
   \\ & \\
     (a)\hspace*{4cm} & \hspace*{-4cm} (b)
    \end{tabular}
    \captions{Gravitino BR to three-body decays as function of the gravitino mass for several low-energy values of $M_1$ and
(a) $M_2 = 2 \, M_1$, (b) $M_2 = 1.1 \, M_1$. In both cases, 
 the following representative values of the low-energy parameters are used: $\lambda = 0.1 $, $\kappa = 0.1$, $\tan\beta = 10$, $v_{\nu^c}$ = 1750 GeV. 
}
    \label{fig2}
\end{figure}

Once these strategies allow us to suppress the photino composition of the neutrinos, 
the BR to $\gamma\nu$ is also suppressed, and the BR 
to three-bodies becomes more important, as mentioned above. 
We show this behavior in Fig.~\ref{fig2} for two different relations between $M_1$ and $M_2$. In panel (a), the BR for three-body final states versus the gravitino mass is shown for several values of $M_1$ at low energy, assuming the approximate grand unified theory (GUT) relation $M_2=2M_1$. For the other parameters we use typical values 
$\lambda = 0.1 $, $\kappa = 0.1$, $\tan\beta = 10$, $v_{\nu^c}$ = 1750 GeV. Variations in these values do not modify our analysis significantly.
As we can see, the three-body final states can give an important contribution to the decay of the gravitino in several ranges of gravitino and gaugino masses, specially for $M_1\gsim 1$ TeV.
In panel (b), we show the same cases as in panel (a) but now using the low-energy relation
$M_2=1.1M_1$. This example is close to the limiting case discussed above, $M_2 \to M_1$, in order to get a cancellation of the mixing parameter. The results in the figure confirm our discussion, and we can see that the three-body final states are even more important than in panel (a). In particular,
already for $M_1=-200$ GeV this BR is larger than 0.5 when $m_{3/2}\gsim 17$ GeV, and
for example, for $M_1=-1$ TeV this is obtained when $m_{3/2}\gsim 4$ GeV.

In order to compute numerically the range of the photino-neutrino mixing parameter, where the 
correct neutrino experimental pattern as presented 
in \cite{Tortola:2012te,Fogli:2012ua,Capozzi:2013csa,Forero:2014bxa,Gonzalez-Garcia:2015qrr} within a three sigma range is reproduced,
we have performed the following scan of the low-energy parameter space of the $\mn$:
\begin{center}
   \begin{tabular}{ r  c  l }
	$0.1 \leq$ & $\lambda$ & $\leq 0.4$, \cr
	$0.1 \leq$ & $\kappa$ & $\leq 0.55$, \cr
	$5 \leq$ & $\tan\beta$ & $\leq 30$, \cr
	$10^{-6}\  \text{GeV} \leq$ & $v_{\nu_{1}}$ & $\leq 10^{-3}\ \text{GeV}$, \cr
	$10^{-6}\ \text{GeV}\leq$ & $v_{\nu_{2,3}}$ & $\leq 10^{-4}\ \text{GeV}$, \cr
	$500\ \text{GeV} \leq$ & $v_{\nu^{c}}$ & $\leq 5\ \text{TeV}$, \cr
	$10^{-8} \leq$ & $Y_{\nu_{1}}$ & $\leq 10^{-6}$, \cr
	$10^{-7} \leq$ & $Y_{\nu_{2,3}}$ & $\leq 10^{-5}$, \cr
	$-75\ \text{TeV} \leq$ & $M_{1}$ & $\leq -200\ \text{GeV}$, \cr
	$0.5 \leq$ & $M_{2}/M_{1}$ & $\leq 2$, \cr
	\end{tabular}
\end{center}
where we use diagonal Yukawa couplings following the discussion in Section~\ref{model}, and the limits on gaugino masses discussed above are taken into account.
In order to find solutions allowed by experimental results on neutrino physics in the easiest possible way we choose negative values for the gaugino masses as discussed in Ref.~\cite{neutrinocp}. 
Thus we
obtain the following result:
\bea
10^{-20} \lesssim|U_{\tilde{\gamma} \nu}|^2\lesssim 10^{-14}\ ,
\label{new_range}
\eea
which extends the lower bound of the previous estimation (\ref{relaxing}).
This lower bound is achieved using the limiting cases for some parameters such as e.g. $M_1=-75$ TeV, $M_2/M_1=1$. 
In this work, we do not need to include such extreme values of gaugino masses in the analysis, since the results do not change significantly with respect to the ones used here,
$|M_1|\leq 10$ TeV. 
Notice also that the results are not going to be essentially modified if we allow the possibility of CP violation in the model.
Complex VEVs in Eq. (\ref{UAP1}) are not not going to change the value of the decay width in Eq. (\ref{decay2body}), since it is proportional to the modulus of the mixing parameter. Therefore, no modification is expected for the flux in the case of the line. Concerning the three-body decays, as can be seen from the Appendix, unless a cancellation between different terms is imposed, the flux is not going to change significantly. We have checked this numerically.





Given our conclusions concerning the range of $|U_{\tilde{\gamma} \nu_i}|$, as well as the three-body final state contributions, an extension of the analysis of the constraints on the $\mn$ parameter space from
\Fermi LAT data, including the new limits, is of great importance.
The next sections are focused on this analysis.


\section{Gamma-ray flux from gravitino decay}
\label{sec:gamma}

The contribution from
gravitino DM decay to the $\gamma$-ray emission observed by 
\Fermi LAT
can have three different sources: i) the smooth Galactic halo, ii) sub-halos hosted by the Galactic halo, iii) extragalactic structures. The signal from extragalactic gravitino decay is expected to be isotropic and, unlike the annihilation case,
independent on the amount of DM clustering at each given 
redshift~\cite{2012MNRAS.421L..87S,2002PhR...372....1C,Sefusatti:2014vha}. 
We will consider the emission from i) and iii) in the analysis. This is the most conservative approach and also the less model dependent. Thus   
\begin{equation}\label{eq:decayFluxTotal}
 \frac{d\Phi_{\gamma}^{\text{total}}}{dEd\Omega}= \frac{d\Phi_{\gamma}^{\text{halo}}}{dEd\Omega} +  \frac{d\Phi_{\gamma}^{\text{extragal}}}{dEd\Omega}\ .
\end{equation}

The differential flux of $\gamma$-rays from gravitino decay in the Galactic halo is calculated by integrating the DM distribution around us along the line of sight: 
%
\begin{equation}\label{eq:decayFlux}
 \frac{d\Phi_{\gamma}^{\text{halo}}}{dEd\Omega}=\frac{1}{4\,\pi\,\tau_{3/2}\,m_{3/2}}\,\frac{1}{\Delta\Omega}\,\frac{dN^{\text{total}}_{\gamma}}{dE} \int_{\Delta\Omega}\!\!\cos
 b\,db\,d\ell\int_0^{\infty}\!\! ds\,\rho_{\text{halo}}(r(s,\,b,\,\ell))\ ,
\end{equation}
where $b$ and $\ell$ denote the Galactic latitude and longitude,
respectively, and $s$ denotes the distance from the Solar System. Furthermore, $\Delta \Omega$ is the ROI.
The radius $r$ in the DM halo density profile of the Milky Way, $\rho_{\text{halo}}$, is expressed in terms of these Galactic coordinates as
\begin{equation}
 r(s,\,b,\,\ell)=\sqrt{s^2+R_{\odot}^2-2\,s\,R_{\odot}\cos{b}\cos{\ell}}\ ,
\end{equation}
where $R_{\odot}\simeq 8.5$ kpc is the radius of the solar orbit around the Galactic center. 
The total number of photons produced in gravitino decay can be expressed as 
\begin{equation}
\frac{dN_{\gamma}^{\text{total}}}{dE}=\sum_{i} BR_i\frac{dN_{i}}{dE}\ ,
\label{eq:dndephotona}
\end{equation}
\noindent where $dN_{i}/dE$ is the photon energy spectrum produced by the different gravitino decay channels studied in Section \ref{sec:Scan}. To compute $dN_{i}/dE$  with Pythia 8.205 \cite{Sjostrand:2014zea} we have created a custom resonance with an energy equal to the gravitino mass that only allows it to decay into a particular channel $i$. 
Then Pythia hadronizes the products and decays the hadrons mainly into leptons that lead to photons through QED processes. The events are stored in a histogram, from which we create a lookup table involving all the possible gravitino decay channels for a set of masses. To accomplish this we use the Monash tune \cite{Skands:2014pea} to run Pythia.

On the other hand,
the prompt contribution from extragalactic structures can be modeled as:
\begin{equation}\label{eq:decayFlux2}
 \frac{d\Phi_{\gamma}^{\text{extragal}}}{dEd\Omega}=\frac{c}{4\pi}\frac{\Omega_{DM}\rho_{c}}{m_{3/2}\tau_{3/2}}\frac{E_{\gamma}}{H_0}\int^{\infty}_{E_{\gamma}}{dE'_{\gamma}\frac{E_{\gamma}}{E'_{\gamma}}\frac{Q_{\gamma}(E_{\gamma},E'_{\gamma})}{\sqrt{\Omega_{\Lambda}+\Omega_M(E_{\gamma},E'_{\gamma})^{3}}}}\ ,
\end{equation}
where $c$ is the speed of light, and we use the values of the cosmological parameters from Planck Collaboration combined with WMAP~\cite{Ade:2013zuv}: 
$H_0 = 67.04$ km s$^{-1}$ Mpc$^{-1}$, 
$\Omega_M = 0.3183$, $\Omega_{DM} = 0.2678$, $\Omega_{\Lambda} = 0.6817$, and $\rho_c = 1.054\times 10^{-5}$ h$^2$ GeV cm$^{-3}$.
Besides,
$E'_{\gamma}=(1+z)E_{\gamma}$ is the energy of $\gamma$ rays when they are produced at redshift $z$, and
\begin{equation}\label{eq:Q}
 Q_{\gamma}(E_{\gamma},E'_{\gamma})=e^{-\tau(z,E_{\gamma})}(1+z)\frac{dN_{\gamma}^{\text{total}}}{dE}
 \ .
\end{equation}
\noindent In this expression $\tau(z,E_{\gamma})$ is the optical depth, for which we adopt the result given in Ref.~\cite{2012MNRAS.422.3189G}. 

Using these formulas, we will compute in Section~\ref{sec:results} the spectral shape and the flux expected from decaying gravitino DM in the $\mn$.
As our theoretical predictions must be compared with the $\gamma$-ray observations, in the next subsection we will discuss the observations of the 
$\gamma$-ray sky by \Fermi LAT that are relevant for our computations.  

\subsection{\Fermi LAT observations}

The $\gamma$-ray sky has been observed by \Fermi LAT with unprecedented detail. Most of the $\gamma$ rays detected come from point-like or small extended sources, and a strong diffuse emission correlated with Galactic structures~\cite{diffuse2}. In addition, a tenuous  diffuse component has been detected, the isotropic $\gamma$-ray background (IGRB)~\cite{2010PhRvL.104j1101A}. The origin of the IGRB can be sources that remain below the detection threshold of \Fermi LAT, among others. For instance, DM decay/annihilation~\cite{Bergstrom:2001jj,Ullio:2002pj} can produce a sizable contribution to the IGRB\footnote{The main IGRB contributors are blazars, star-forming galaxies, diffuse processes such as intergalactic 
shocks~\cite{Colafrancesco:1998us,Loeb:2000na,Zandanel:2013wea}, interactions of ultra high energy cosmic rays with the extragalactic background light (EBL)~\cite{Berezinsky:1975zz}, and cosmic-ray interactions in small solar-system bodies~\cite{Moskalenko:2009tv}.}. The observed IGRB depends on the point source detection threshold of the instrument. Instead, a physical quantity is the total EGB, defined as the combination of resolved sources and the IGRB. The \Fermi LAT Collaboration has determined the EGB using 50 months of data reprocessed with the Pass 8 event-level analysis that expands from 100 MeV to 820 GeV~\cite{Ackermann:2014usa}.
In Fig.~3 of that work, the integrated LAT counts above 100 MeV that are used in the analysis are shown, as well as the regions in the vicinity of the Galactic plane that have been masked. 
Recently, the authors of Ref. \citep{Baring:2015sza} have analysed constraints on two-body dark matter decays using {\it Fermi}-LAT gamma-ray data from the observation of dwarf spheroidal galaxies. Although this kind of analysis, applied to three-body decays, could also be of interest in our context, it is clearly beyond the scope of this paper, and we leave it for future studies.

\begin{figure}[t!]
 \begin{tabular}{cc}
 \hspace*{-4mm}
       \epsfig{file=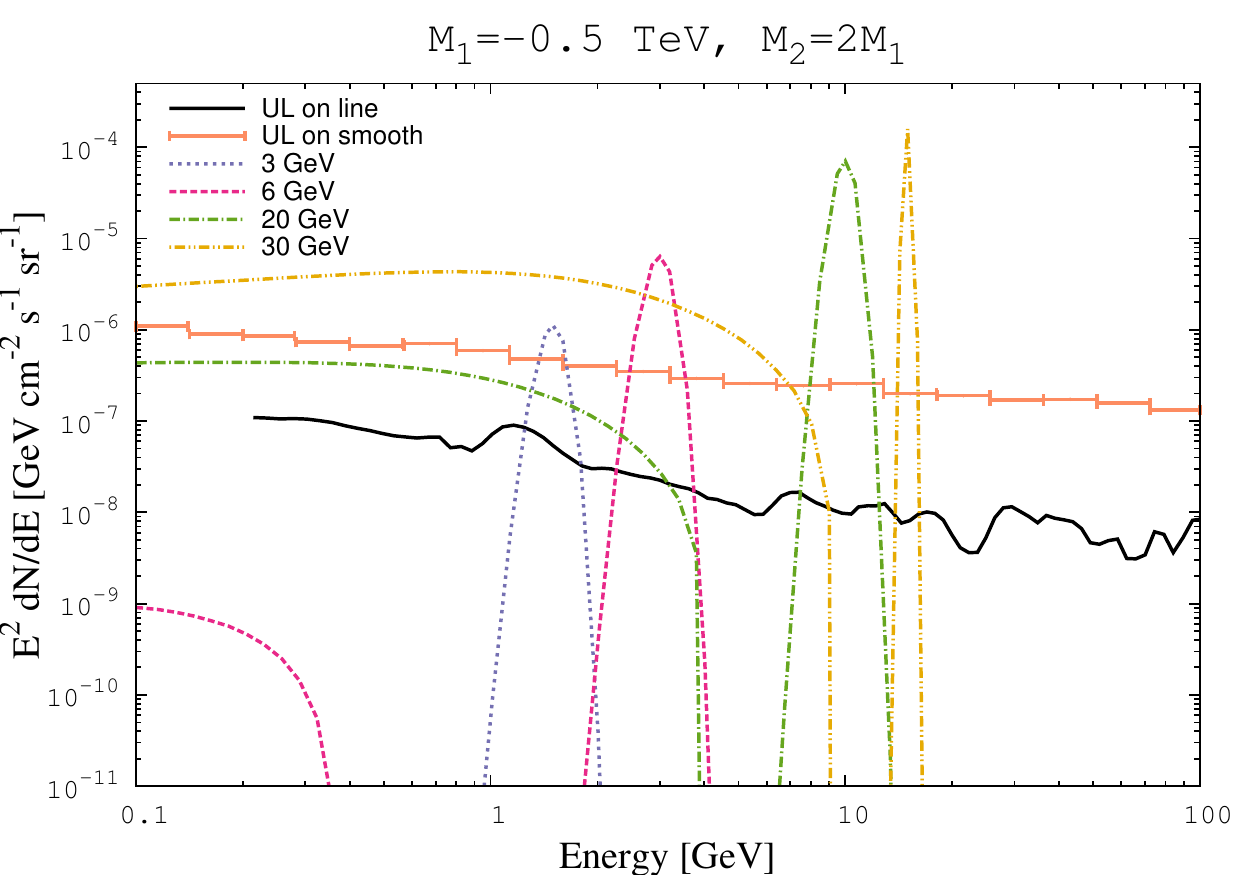,height=5.4cm}
       \epsfig{file=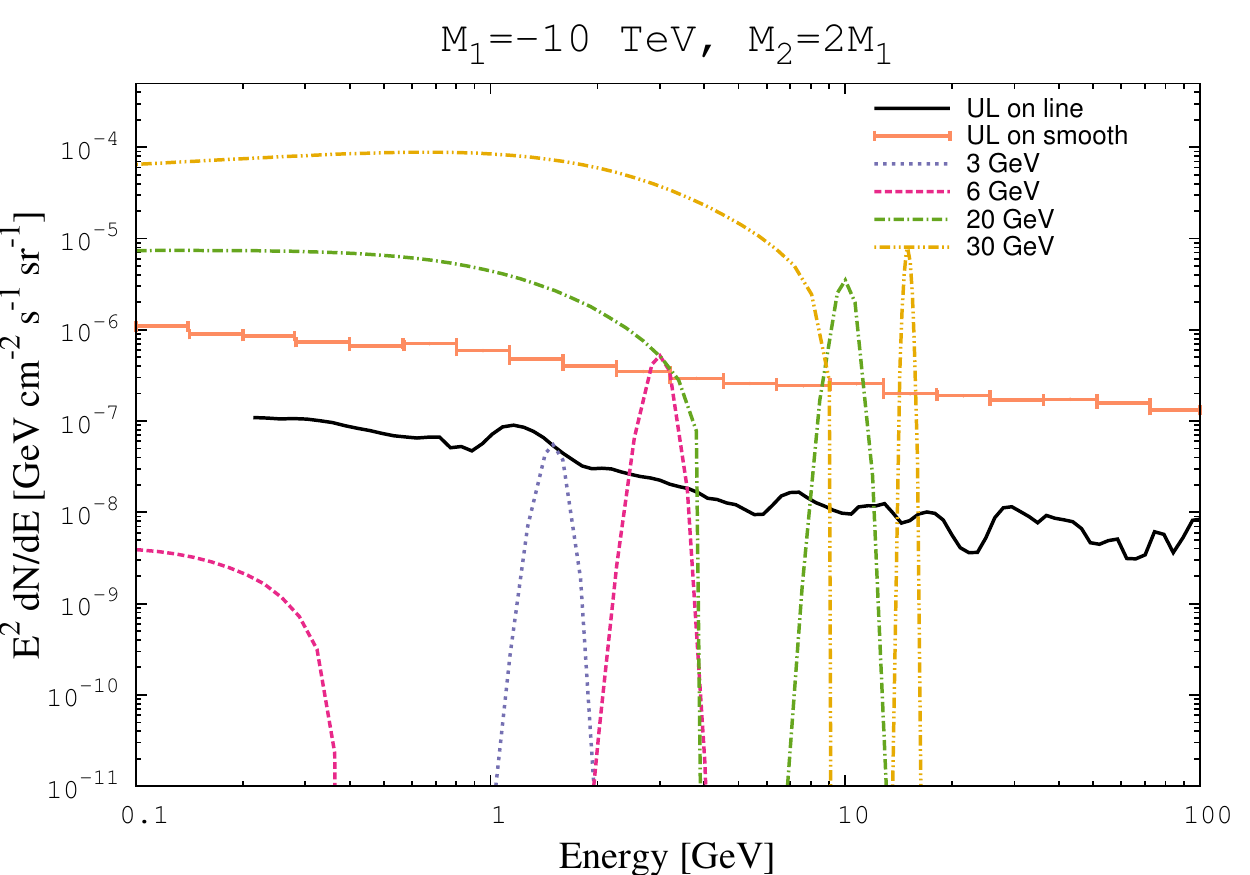,height=5.4cm} 
       \vspace*{-0.8cm}       
   \\ & \\
          (a)\hspace*{4cm} & \hspace*{-4cm} (b)
    \\ & \\
\hspace*{-4mm}
       \epsfig{file=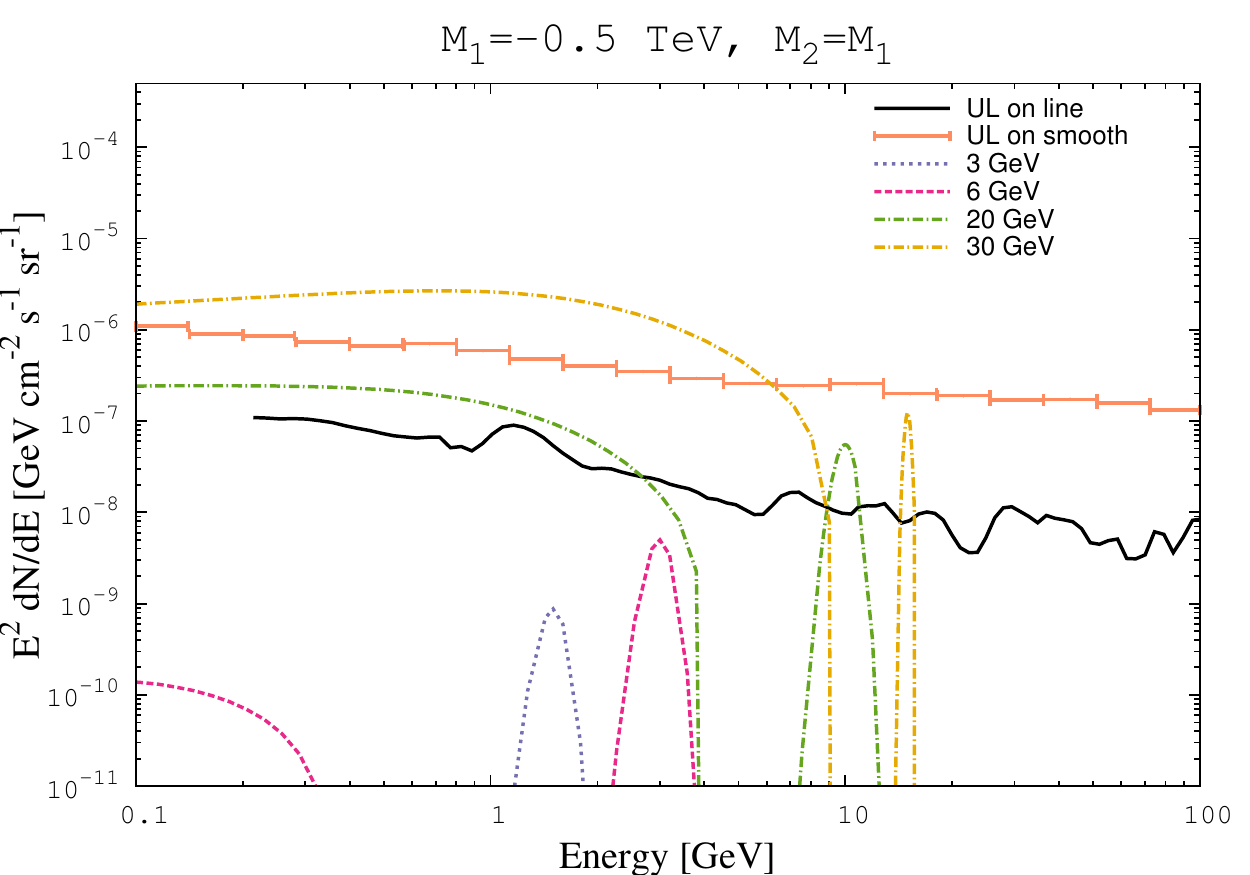,height=5.4cm}
       \epsfig{file=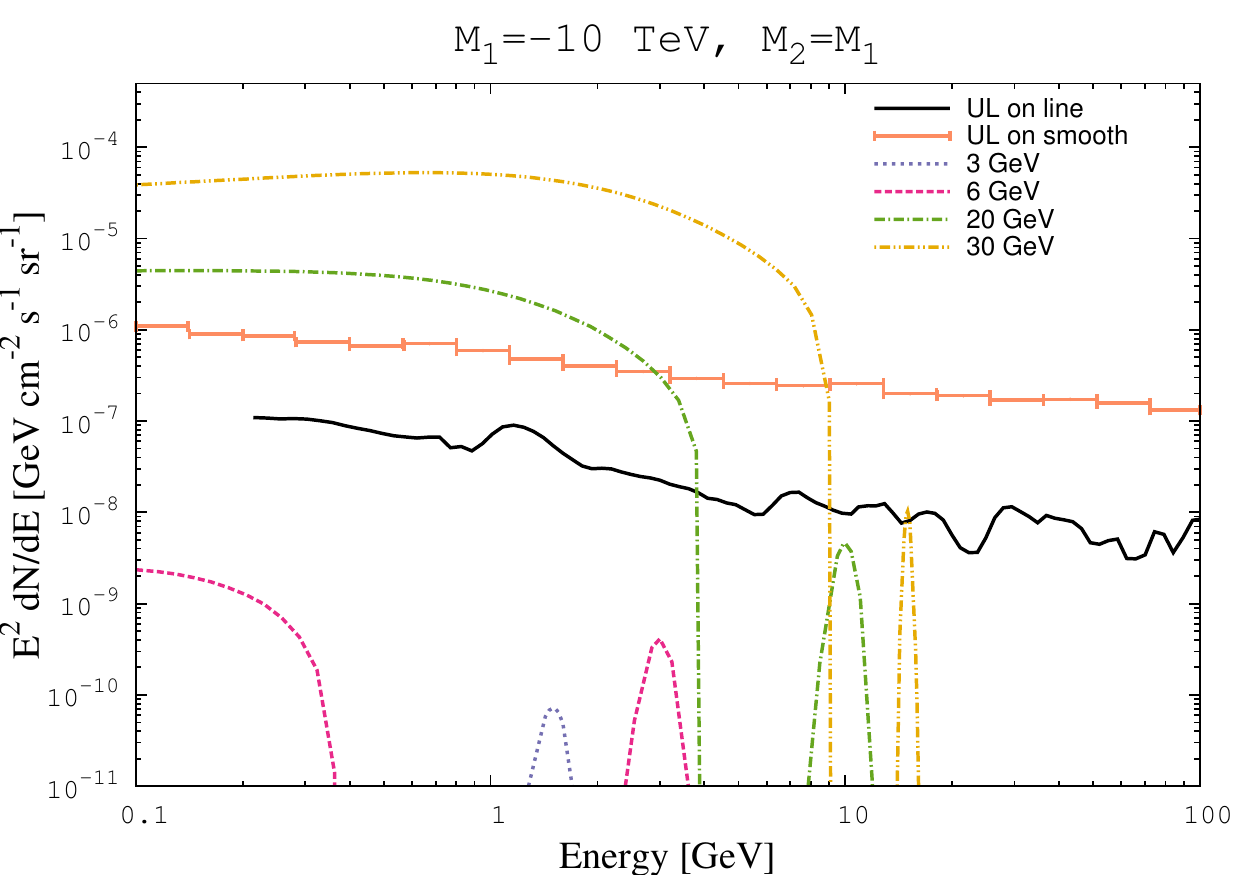,height=5.4cm} 
       \vspace*{-0.8cm}       
   \\ & \\
         (c)\hspace*{4cm} & \hspace*{-4cm} (d)
    \end{tabular}
\captions{Expected $\gamma$-ray spectrum for several examples of gravitino DM decay,
$m_{3/2}=3,6,20,30$ GeV, using four limiting combinations of 
gaugino masses, (a) $M_2=2M_1$ with $M_1=-0.5$ TeV, (b) $M_2=2M_1$ with $M_1=-10$ TeV, 
(c) $M_2=M_1$ with $M_1=-0.5$ TeV, (d) $M_2=M_1$ with $M_1=-10$ TeV.
The black line corresponds to the upper limits (UL) from line searches, whereas the orange points
corresponds to the upper limits from EGB determination after subtracting models of known contributors.
For both limits, \Fermi LAT data are used.
}
    \label{fig_EGBa}
\end{figure}

In Fig.~\ref{fig_EGBa} we show the 95\% CL upper limit from EGB determination (orange points). 
In order to get this limit the average emission from non-exotic contributors are subtracted\footnote{We assume Gaussian errors, therefore 95\% of the area of a Gaussian distribution is within 1.64 standard deviations of the mean.}. The non-exotic contributors to the EGB considered are: star-forming galaxies~\cite{Ackermann:2012vca}, radio galaxies~\cite{Inoue:2011bm} and the integrated emission of blazars with EBL absorption as recently modeled in~\cite{AjelloBlazars}. The limits are taken from~\cite{Carquin:2015uma}. We use these limits to probe the smooth spectral gravitino signal.

The sharp spectral feature at half of the gravitino mass is in general the brightest feature on the
gravitino-induced $\gamma$-ray spectrum. We use limits on line emission (black line) from an updated analysis by the \Fermi LAT Collaboration~\cite{Ackermann:2015lka}, where they use the so-called R180 ROI. Such ROI is defined as a circular region of radius 180$^o$ centered on the Galactic center.
In addition, the Galactic plane region with longitude greater than 6$^o$ from the Galactic center and latitude smaller than 5$^o$ is removed (see Fig.~4 of Ref.~\cite{Ackermann:2015lka}).
The limits are set using the analysis methods developed in~\cite{Albert:2014hwa} to account for systematic uncertainties at the low-energy end of the \Fermi LAT band, and 69.9 months of Pass 8 data.

Therefore, in the next section we will use the limits from EGB~\cite{Carquin:2015uma} 
and the latest spectral line search by the \Fermi LAT Collaboration~\cite{Ackermann:2015lka} to set constraints on the parameter space of $\mn$ gravitino dark matter.

\section{Constraints on $\mn$ gravitino dark matter from \Fermi LAT data}
\label{sec:results}

We apply now the formulas of Sections~\ref{sec:Scan} and~\ref{sec:gamma}
to compute the flux of $\gamma$ rays from decaying gravitino DM in the $\mn$,  following 
the analysis of Ref.~\cite{Albert:2014hwa} regarding 
the application of \Fermi LAT exclusion limits.

In Fig.~\ref{fig_EGBa} we plot the spectral shapes for four different gravitino masses, 
$m_{3/2}=3, 6, 20, 30$ GeV. For each case,
the sharp feature at half of the gravitino mass corresponds to the usual spectral line from the two-body decay channel, convolved with the \Fermi LAT energy dispersion\footnote{http://www.slac.stanford.edu/exp/glast/groups/canda/lat\_Performance.htm}. 
As expected from Eqs.~(\ref{decay2body})
and~(\ref{eq:decayFlux}), the larger $m_{3/2}$ the larger is the decay width and the flux, yielding a brighter
sharp feature.
On the other hand, the smooth bump corresponds to the emission from the three-body decay channels. 
To set constraints we compare the sharp spectral feature with the upper limits from lines (black line) and the smooth bump with the upper limits from EGB (orange points), computing
the gravitino signal accordingly to the ROIs used to obtain such limits.
The panels (a) and (b) 
correspond to scenarios with the GUT-inspire relation at low energy, $M_2 = 2 \, M_1$.
As we can see, all the gravitino masses presented are excluded by line limits in both panels, but the
case of 3 GeV in panel (b) where gaugino masses are increased by more than one order of magnitude. This is consistent with results in Ref.~\cite{Albert:2014hwa}. 
However, in panels (c) and (d), where the limiting case for the relation between 
gaugino masses, $M_2 = M_1$, is considered, 
the suppression on the line strength is evident with respect to the previous panels.
As expected, making the values of the gaugino masses similar is more efficient to evade \Fermi LAT limits on lines than increasing their absolute values.
Now, gravitino masses of 3 and 6 GeV are allowed in both panels (c) and (d). 
Although a mass of 20 GeV is allowed by the line limits for the case $M_1=-10$ TeV, at the end of the day EGB limits forbids it. 

Summarizing, an increasing in the gaugino masses $M_{1,2}$ and/or a 
shrinking of their ratio $M_2/M_1$ suppress the line size,
however the line limits are restrictive enough to ruled out almost all scenarios but the extreme ones, i.e. with large and close gaugino masses.

\begin{figure}[t!]
 \begin{center}
       \epsfig{file=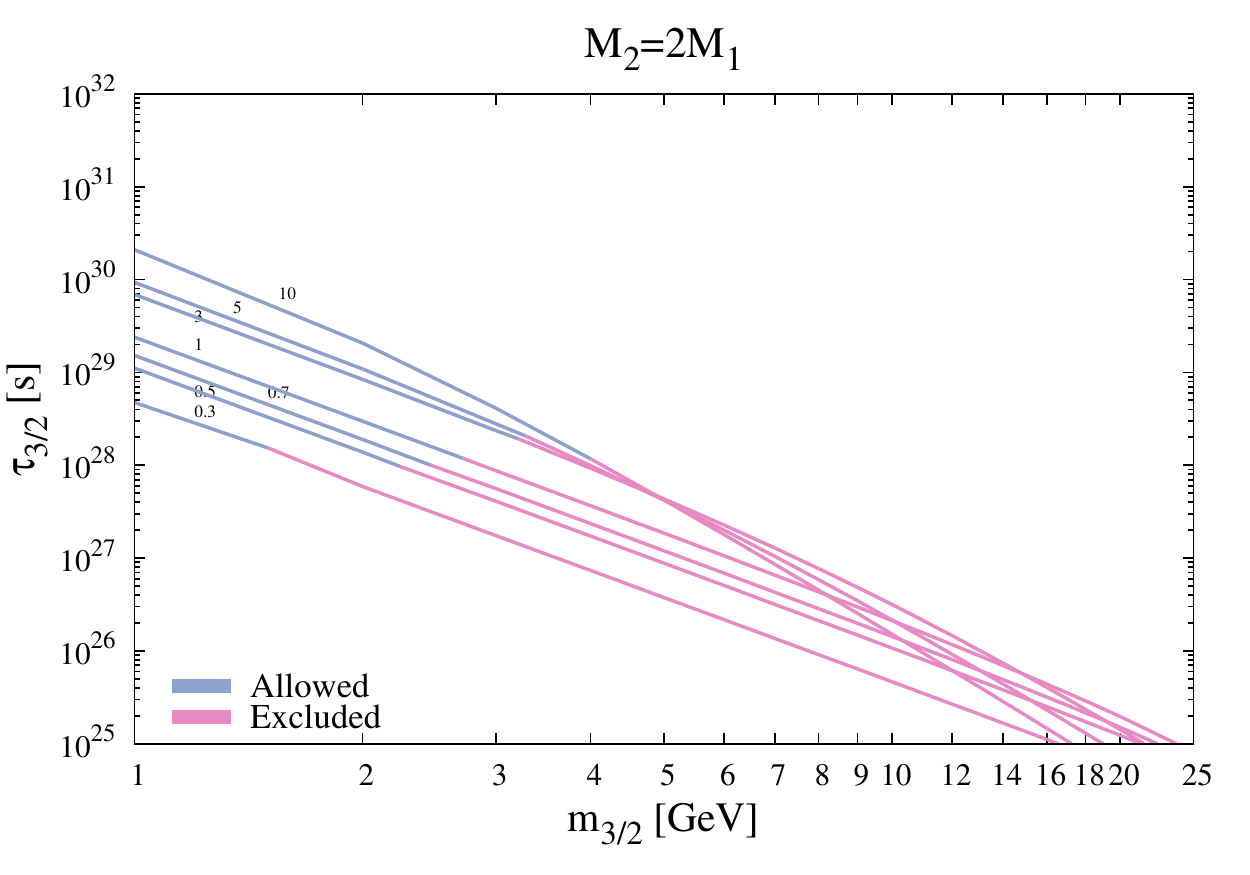,height=6.5cm}
       \epsfig{file=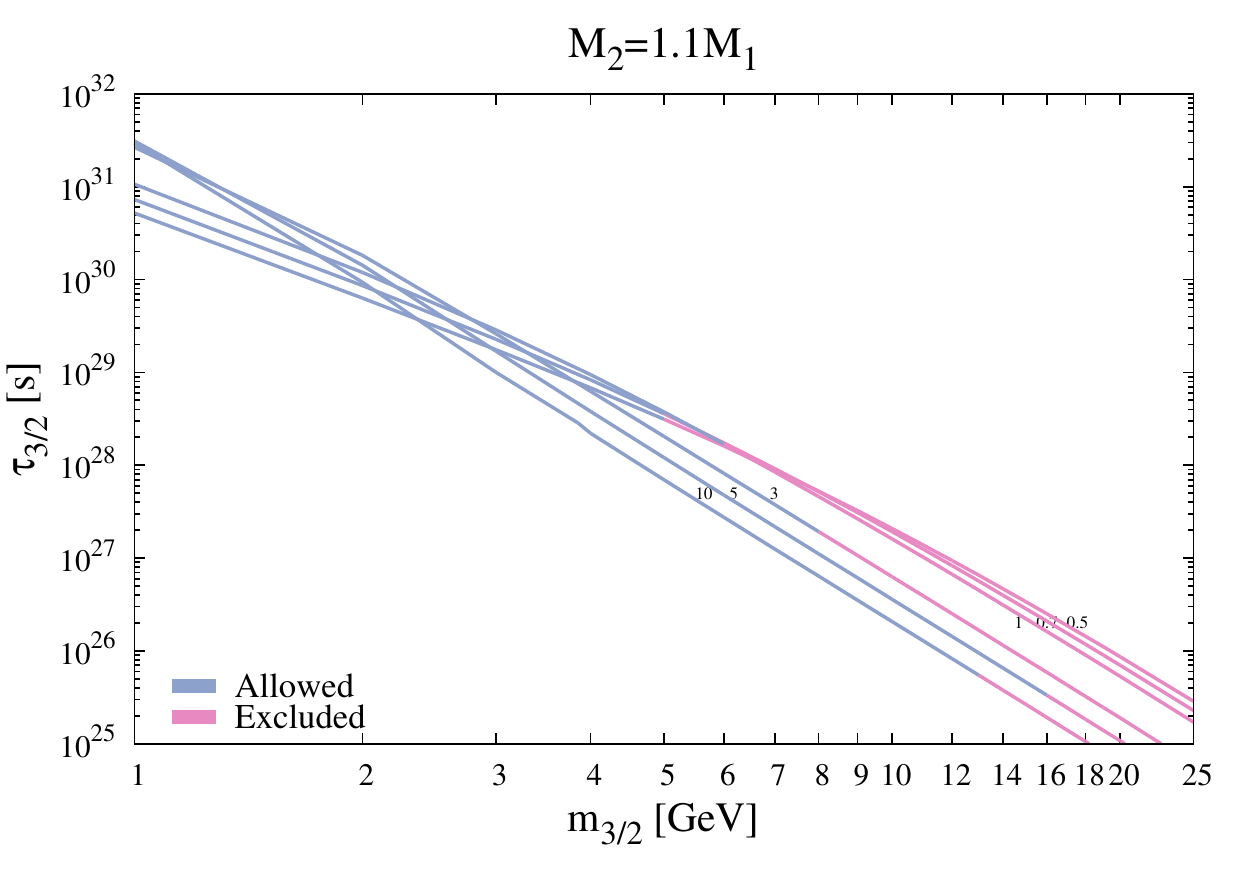,height=6.5cm}
       \epsfig{file=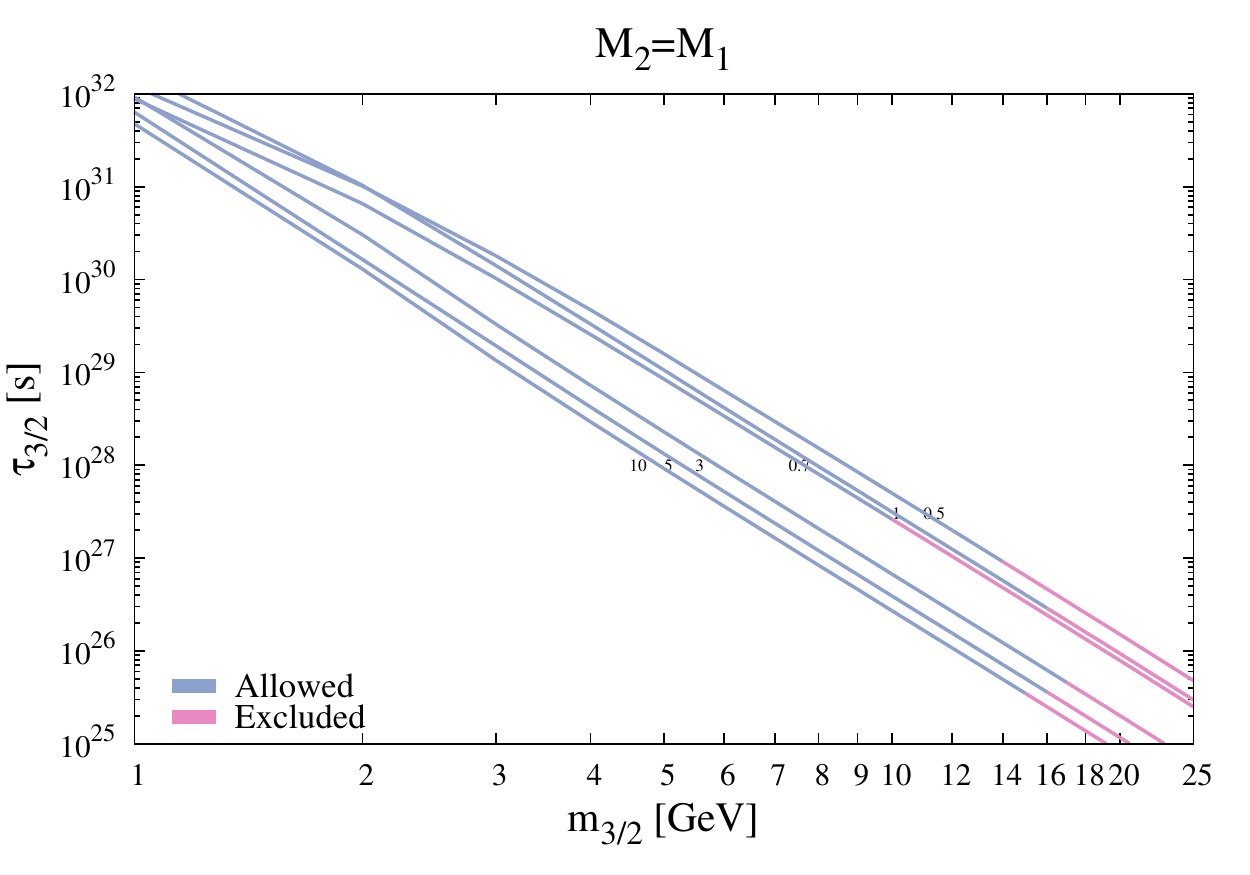,height=6.5cm}
  \captions{Parameter space of decaying gravitino DM given in terms of the gravitino lifetime and the gravitino mass, using three relations for gaugino masses $M_2=2M_1$, $M_2=1.1 M_1$ and $M_2=M_1$. 
For each of the three relations we plot the values $M_1=-0.5, -0.7, -1, -3, -5, -10$ TeV, and label the curves with $|M_1|$.
The blue 
section of each curve is allowed whereas the magenta one is forbidden by \Fermi LAT data.
}
    \label{figconstrains}
\end{center}
\end{figure}

In Fig.~\ref{figconstrains} we show the exclusion regions in the plane gravitino lifetime versus gravitino mass
for three different relations between $M_1$ and $M_2$: $M_2=2M_1$, $M_2=1.1 M_1$ and $M_2=M_1$.
For each of the three cases we plot the following values 
of $M_1$: $M_1=-0.5, -0.7, -1, -3, -5, -10$ TeV.
The same representative values of the low-energy parameters as in Fig.~\ref{fig2} are used:
$\lambda = 0.1 $, $\kappa = 0.1$, $\tan\beta = 10$, $v_{\nu^c}$ = 1750 GeV. 
In Tables~\ref{param2}, \ref{param11} and \ref{param1} we show for each case the relevant parameters in order to reproduce the observed neutrino masses and mixing angles, together with the corresponding values of the photino-neutrino mixing parameter.
There we can see that the latter reaches the lower bound only in the limiting cases with
$M_2 \to M_1$.
The blue section of each curve in Fig.~\ref{figconstrains} indicates the values of $m_{3/2}$ and $\tau_{3/2}$ that are allowed, whereas the magenta one indicates those forbidden.
In general, the dominant exclusion is produced by the line limits coming from the two-body decay channel, because
as discussed above in Fig.~\ref{fig_EGBa} in most of the cases the sharp spectral signal gives the brightest feature, and additionally the \Fermi LAT limits on line flux are stronger than the limits on the smooth spectrum from EGB.

Nevertheless, the smooth signal plays also an important role by changing the lifetime of the gravitino, hence modifying the exclusion limits for its parameter space. See for example in Fig.~\ref{figconstrains} the cases $M_2=1.1 M_1$ and
$M_2= M_1$, where lifetimes as low as about $4\times 10^{25}$ s are allowed.
Another important issue to notice is the fact that the curves are crossing each other. To understand this we need to take into account that the slope of each curve, when the decay is dominated by the two-body final states, is different from the slope when the three-body final states are the dominant ones. The zone in the plane when this change of regime takes place is different for the different curves showed in the figure. Also a significant fact for the analysis is that, as already mentioned, if we take larger values for the gaugino masses the photino-neutrino mixing parameter decreases implying less constrains by \Fermi LAT limits on lines, but after certain value the diffuse spectra dominates completely the slope of the curves. This is the case for $M_1\gsim 10$ TeV which results in an increasing of the exclusion of the parameter space, i.e. lower allowed values for the gravitino mass, as we can see in Fig.~\ref{figconstrains} for the limiting cases $M_2=1.1 M_1$ and $M_2=M_1$.

Summarizing the results, for $M_2=2 M_1$ the line is the dominant feature, and the gravitino mass must be smaller than $4~\gev$ to avoid exclusion by \Fermi LAT data. This is in agreement with Ref.~\cite{Albert:2014hwa}.
For $M_2=1.1M_1$ and $M_1$ below 1 TeV, the exclusion is driven by the two-body decay channel as in the previous case, and still not very different values of $m_{3/2}$ are allowed ($\lsim 6$ GeV). 
However, above 1 TeV the line limits are evaded as the $|U_{\tilde{\gamma} \nu_i}|$ is strongly suppressed by the combination of increasing gaugino masses and making them closer. Gravitino masses as large as
$16~\gev$ can be reached without exclusion.
For $M_2=M_1$, the line is still crucial and even for low value of $M_1$ is possible to reach gravitino masses above $10~\gev$. 



\clearpage 
\clearpage

Therefore, using \Fermi LAT data, we have obtained the following constraints on the gravitino mass and lifetime in the $\mn$:
$m_{3/2}\lsim 17~\gev$ and $\tau_{3/2}\gsim 4\times 10^{25}$ s.

\begin{figure}[t]
 \begin{center}
       \epsfig{file=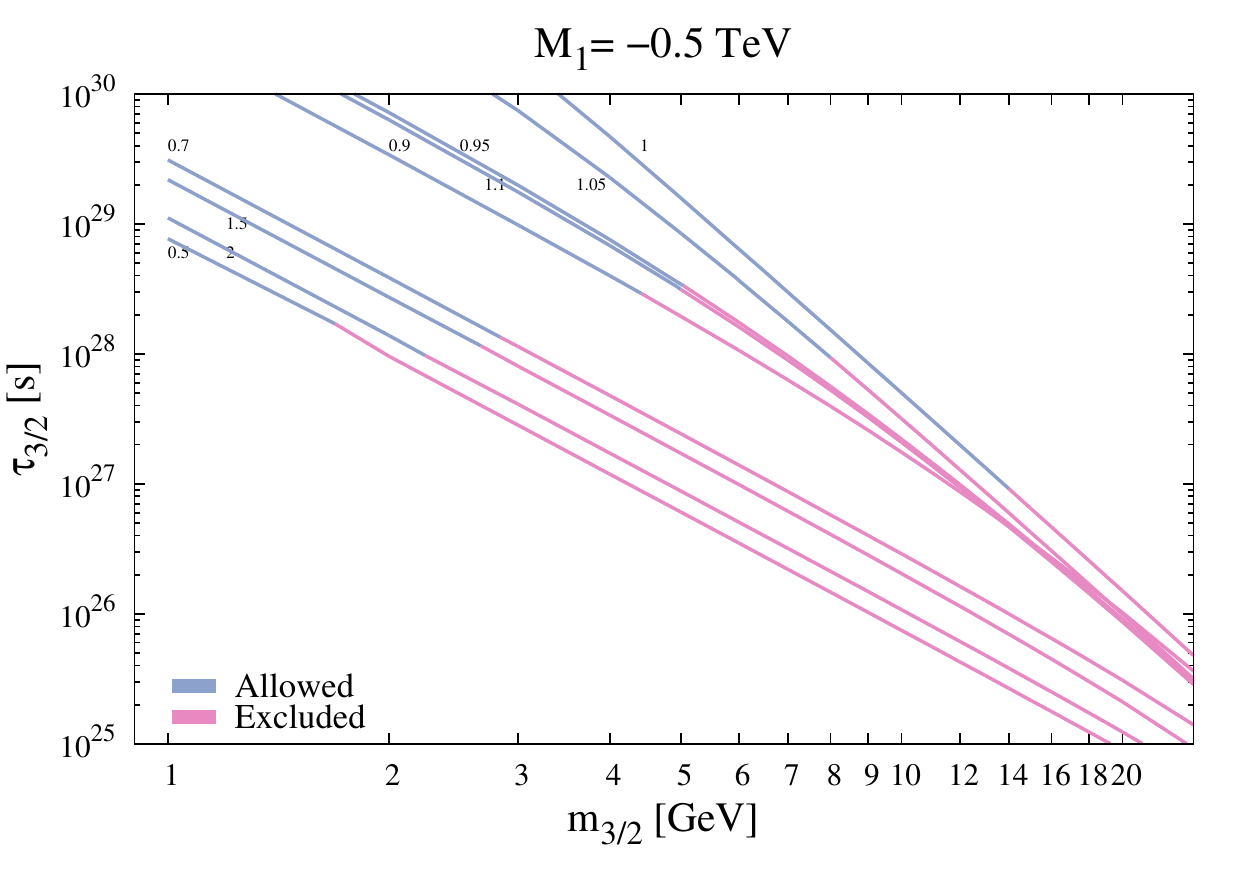,height=6.5cm}
\captions{The same as in Fig.~\ref{figconstrains} but using $M_1=-0.5$ TeV and 
$M_2 = x M_1$ with $x=0.5,0.7,0.9,0.95,1,1.05,1.1,1.5,2$. Each curve is labeled with the corresponding value of $x$.
}
    \label{figsymmetric12}
 \end{center}
\end{figure}

Finally, we might think of the possibility of modifying these constraints using
$M_2 = x M_1$ with $x<1$, in addition to $x>1$ as we have been using so far.
Notice however from the approximation~(\ref{UAP1}) that the spectrum of the two-body decay is proportional to the ratio $|(1-x)/x|$, implying that the larger this ratio the 
stronger are the constraints on $m_{3/2}$. Thus the results for $x<1$ will not be essentially different from those for $x>1$.
This argument is not exact for the spectrum of the three-body decays. For instance in the channels involving $W$ bosons the mixing comes from the chargino-charged lepton matrix, where only $M_2$ appears.
Nevertheless, the numerical analysis shows that the above argument can still be used.
We show this fact in Fig.~\ref{figsymmetric12} 
for $M_1=-0.5 \tev$ and different values of $x$. For example, the strongest (weakest) constraint on the gravitino mass is obtained for $x=0.5$ ($x=1$).

\clearpage
\begin{table}
\begin{center}
   \begin{tabular}{| c | c | c | c | }
    \hline
       $M_1=-0.3 \text{TeV}$  &  $v_{\nu_1}=1.7\times10^{-4} \text{GeV}$  & $v_{\nu_2}=0.2\times10^{-4} \text{GeV}$   &  $v_{\nu_3}=0.8\times10^{-4} \text{GeV}$     \\ \hline
       $|U_{\tilde{\gamma} \nu}|^2=7.85\times10^{-15}$ &  $Y_{\nu_{1}}=5\times10^{-7}$  &  $Y_{\nu_{2}}=0.8\times10^{-6}$  &   $Y_{\nu_{3}}=0.9\times10^{-6}$ \\ \hline \hline
       $M_1=-0.5 \text{TeV}$  &  $v_{\nu_1}=1.5\times10^{-4} \text{GeV}$  & $v_{\nu_2}=0.2\times10^{-4} \text{GeV}$   &  $v_{\nu_3}=1.5\times10^{-4} \text{GeV}$     \\ \hline
       $|U_{\tilde{\gamma} \nu}|^2=3.35\times10^{-15}$ &  $Y_{\nu_{1}}=3\times10^{-7}$  &  $Y_{\nu_{2}}=0.7\times10^{-6}$  &   $Y_{\nu_{3}}=0.9\times10^{-6}$ \\ \hline \hline
       $M_1=-0.7 \text{TeV}$  &  $v_{\nu_1}=1.7\times10^{-4} \text{GeV}$  & $v_{\nu_2}=2\times10^{-4} \text{GeV}$   &  $v_{\nu_3}=0.3\times10^{-4} \text{GeV}$     \\ \hline
       $|U_{\tilde{\gamma} \nu}|^2=2.44\times10^{-15}$ &  $Y_{\nu_{1}}=3\times10^{-7}$  &  $Y_{\nu_{2}}=0.9\times10^{-6}$  &   $Y_{\nu_{3}}=0.7\times10^{-6}$ \\ \hline \hline
       $M_1=- 1\text{TeV}$  &  $v_{\nu_1}=2.3\times10^{-4} \text{GeV}$  & $v_{\nu_2}=2\times10^{-4} \text{GeV}$   &  $v_{\nu_3}=0.6\times10^{-4} \text{GeV}$     \\ \hline
       $|U_{\tilde{\gamma} \nu}|^2=1.55\times10^{-15}$ &  $Y_{\nu_{1}}=3\times10^{-7}$  &  $Y_{\nu_{2}}=0.9\times10^{-6}$  &   $Y_{\nu_{3}}=0.7\times10^{-6}$ \\ \hline \hline
       $M_1=-3 \text{TeV}$  &  $v_{\nu_1}=4.3\times10^{-4} \text{GeV}$  & $v_{\nu_2}=4\times10^{-4} \text{GeV}$   &  $v_{\nu_3}=0.8\times10^{-4} \text{GeV}$     \\ \hline
       $|U_{\tilde{\gamma} \nu}|^2=5.35\times10^{-16}$ &  $Y_{\nu_{1}}=3\times10^{-7}$  &  $Y_{\nu_{2}}=0.9\times10^{-6}$  &   $Y_{\nu_{3}}=0.7\times10^{-6}$ \\ \hline \hline
       $M_1=-5 \text{TeV}$  &  $v_{\nu_1}=6.1\times10^{-4} \text{GeV}$  & $v_{\nu_2}=6.2\times10^{-4} \text{GeV}$   &  $v_{\nu_3}=0.4\times10^{-4} \text{GeV}$     \\ \hline
       $|U_{\tilde{\gamma} \nu}|^2=3.91\times10^{-16}$ &  $Y_{\nu_{1}}=3\times10^{-7}$  &  $Y_{\nu_{2}}=0.9\times10^{-6}$  &   $Y_{\nu_{3}}=0.8\times10^{-6}$ \\ \hline \hline
       $M_1=-10 \text{TeV}$  &  $v_{\nu_1}=8\times10^{-4} \text{GeV}$  & $v_{\nu_2}=8.1\times10^{-4} \text{GeV}$   &  $v_{\nu_3}=1.4\times10^{-4} \text{GeV}$     \\ \hline
        $|U_{\tilde{\gamma} \nu}|^2=1.65\times10^{-16}$ &  $Y_{\nu_{1}}=3\times10^{-7}$  &  $Y_{\nu_{2}}=0.9\times10^{-6}$  &   $Y_{\nu_{3}}=0.7\times10^{-6}$ \\ \hline
   \end{tabular}
    \caption{Relevant parameters in order reproduce the observed neutrino masses and mixing angles for the cases
with $M_2=2M_1$ in Fig.~\ref{figconstrains}. 
The corresponding values of the photino-neutrino mixing parameter $|U_{\tilde{\gamma} \nu}|^2$ are also shown.
}
\label{param2} 
\end{center}
\end{table}

\begin{table}
\begin{center}
   \begin{tabular}{| c | c | c | c | }
    \hline
       $M_1=-0.3 \text{TeV}$  &  $v_{\nu_1}=1.3\times10^{-4} \text{GeV}$  & $v_{\nu_2}=0.6\times10^{-4} \text{GeV}$   &  $v_{\nu_3}=0.09\times10^{-4} \text{GeV}$     \\ \hline
       $|U_{\tilde{\gamma} \nu}|^2=1.25\times10^{-16}$ &  $Y_{\nu_{1}}=5\times10^{-7}$  &  $Y_{\nu_{2}}=0.9\times10^{-6}$  &   $Y_{\nu_{3}}=0.8\times10^{-6}$ \\ \hline \hline
       $M_1=-0.5 \text{TeV}$  &  $v_{\nu_1}=1.9\times10^{-4} \text{GeV}$  & $v_{\nu_2}=0.8\times10^{-4} \text{GeV}$   &  $v_{\nu_3}=0.03\times10^{-4} \text{GeV}$     \\ \hline
       $|U_{\tilde{\gamma} \nu}|^2=7.10\times10^{-17}$ &  $Y_{\nu_{1}}=5\times10^{-7}$  &  $Y_{\nu_{2}}=0.9\times10^{-6}$  &   $Y_{\nu_{3}}=0.8\times10^{-6}$ \\ \hline \hline
       $M_1=-0.7 \text{TeV}$  &  $v_{\nu_1}=2.2\times10^{-4} \text{GeV}$  & $v_{\nu_2}=1.2\times10^{-4} \text{GeV}$   &  $v_{\nu_3}=0.2\times10^{-4} \text{GeV}$     \\ \hline
       $|U_{\tilde{\gamma} \nu}|^2=5.05\times10^{-17}$ &  $Y_{\nu_{1}}=5\times10^{-7}$  &  $Y_{\nu_{2}}=0.9\times10^{-6}$  &   $Y_{\nu_{3}}=0.8\times10^{-6}$ \\ \hline \hline
       $M_1=-1 \text{TeV}$  &  $v_{\nu_1}=2.8\times10^{-4} \text{GeV}$  & $v_{\nu_2}=1.3\times10^{-4} \text{GeV}$   &  $v_{\nu_3}=0.3\times10^{-4} \text{GeV}$     \\ \hline
       $|U_{\tilde{\gamma} \nu}|^2=3.40\times10^{-17}$ &  $Y_{\nu_{1}}=5\times10^{-7}$  &  $Y_{\nu_{2}}=0.9\times10^{-6}$  &   $Y_{\nu_{3}}=0.8\times10^{-6}$ \\ \hline \hline
       $M_1=-3 \text{TeV}$  &  $v_{\nu_1}=4.7\times10^{-4} \text{GeV}$  & $v_{\nu_2}=3\times10^{-4} \text{GeV}$   &  $v_{\nu_3}=1.3\times10^{-4} \text{GeV}$     \\ \hline
       $|U_{\tilde{\gamma} \nu}|^2=1.09\times10^{-17}$ &  $Y_{\nu_{1}}=5\times10^{-7}$  &  $Y_{\nu_{2}}=0.9\times10^{-6}$  &   $Y_{\nu_{3}}=0.8\times10^{-6}$ \\ \hline \hline
       $M_1=-5 \text{TeV}$  &  $v_{\nu_1}=6.3\times10^{-4} \text{GeV}$  & $v_{\nu_2}=4\times10^{-4} \text{GeV}$   &  $v_{\nu_3}=1.5\times10^{-4} \text{GeV}$     \\ \hline
       $|U_{\tilde{\gamma} \nu}|^2=6.50\times10^{-18}$ &  $Y_{\nu_{1}}=5\times10^{-7}$  &  $Y_{\nu_{2}}=0.9\times10^{-6}$  &   $Y_{\nu_{3}}=0.8\times10^{-6}$ \\ \hline \hline
       $M_1=-10 \text{TeV}$  &  $v_{\nu_1}=7.2\times10^{-4} \text{GeV}$  & $v_{\nu_2}=7.1\times10^{-4} \text{GeV}$   &  $v_{\nu_3}=0.4\times10^{-4} \text{GeV}$     \\ \hline
       $|U_{\tilde{\gamma} \nu}|^2=2.67\times10^{-18}$ &  $Y_{\nu_{1}}=3\times10^{-7}$  &  $Y_{\nu_{2}}=0.9\times10^{-6}$  &   $Y_{\nu_{3}}=0.8\times10^{-6}$ \\ \hline
   \end{tabular}
    \caption{The same as in Table~\ref{param2} but for $M_2=1.1M_1$.
}
\label{param11} 
\end{center}
\end{table}

\begin{table}
\begin{center}
   \begin{tabular}{| c | c | c | c | }
    \hline
       $M_1=-0.3 \text{TeV}$  &  $v_{\nu_1}=1.3\times10^{-4} \text{GeV}$  & $v_{\nu_2}=0.5\times10^{-4} \text{GeV}$   &  $v_{\nu_3}=0.01\times10^{-4} \text{GeV}$     \\ \hline
       $|U_{\tilde{\gamma} \nu}|^2=6.05\times10^{-18}$ &  $Y_{\nu_{1}}=5\times10^{-7}$  &  $Y_{\nu_{2}}=0.9\times10^{-6}$  &   $Y_{\nu_{3}}=0.8\times10^{-6}$ \\ \hline \hline
       $M_1=-0.5 \text{TeV}$  &  $v_{\nu_1}=1.3\times10^{-4} \text{GeV}$  & $v_{\nu_2}=0.8\times10^{-4} \text{GeV}$   &  $v_{\nu_3}=0.06\times10^{-4} \text{GeV}$     \\ \hline
       $|U_{\tilde{\gamma} \nu}|^2=2.62\times10^{-18}$ &  $Y_{\nu_{1}}=3\times10^{-7}$  &  $Y_{\nu_{2}}=0.9\times10^{-6}$  &   $Y_{\nu_{3}}=0.7\times10^{-6}$ \\ \hline \hline
       $M_1=-0.7 \text{TeV}$  &  $v_{\nu_1}=2.2\times10^{-4} \text{GeV}$  & $v_{\nu_2}=1\times10^{-4} \text{GeV}$   &  $v_{\nu_3}=0.1\times10^{-4} \text{GeV}$     \\ \hline
       $|U_{\tilde{\gamma} \nu}|^2=3.35\times10^{-18}$ &  $Y_{\nu_{1}}=5\times10^{-7}$  &  $Y_{\nu_{2}}=0.9\times10^{-6}$  &   $Y_{\nu_{3}}=0.8\times10^{-6}$ \\ \hline \hline
       $M_1=-1 \text{TeV}$  &  $v_{\nu_1}=1.8\times10^{-4} \text{GeV}$  & $v_{\nu_2}=1.3\times10^{-4} \text{GeV}$   &  $v_{\nu_3}=0.5\times10^{-4} \text{GeV}$     \\ \hline
       $|U_{\tilde{\gamma} \nu}|^2=1.45\times10^{-18}$ &  $Y_{\nu_{1}}=3\times10^{-7}$  &  $Y_{\nu_{2}}=0.9\times10^{-6}$  &   $Y_{\nu_{3}}=0.7\times10^{-6}$ \\ \hline \hline
       $M_1=-3 \text{TeV}$  &  $v_{\nu_1}=4.7\times10^{-4} \text{GeV}$  & $v_{\nu_2}=2.8\times10^{-4} \text{GeV}$   &  $v_{\nu_3}=0.9\times10^{-4} \text{GeV}$     \\ \hline
       $|U_{\tilde{\gamma} \nu}|^2=9.58\times10^{-19}$ &  $Y_{\nu_{1}}=5\times10^{-7}$  &  $Y_{\nu_{2}}=0.9\times10^{-6}$  &   $Y_{\nu_{3}}=0.8\times10^{-6}$ \\ \hline \hline
       $M_1=-5 \text{TeV}$  &  $v_{\nu_1}=6.1\times10^{-4} \text{GeV}$  & $v_{\nu_2}=4\times10^{-4} \text{GeV}$   &  $v_{\nu_3}=1.1\times10^{-4} \text{GeV}$     \\ \hline
       $|U_{\tilde{\gamma} \nu}|^2=6.11\times10^{-19}$ &  $Y_{\nu_{1}}=5\times10^{-7}$  &  $Y_{\nu_{2}}=0.9\times10^{-6}$  &   $Y_{\nu_{3}}=0.8\times10^{-6}$ \\ \hline \hline
       $M_1=-10 \text{TeV}$  &  $v_{\nu_1}=6.2\times10^{-4} \text{GeV}$  & $v_{\nu_2}=6.1\times10^{-4} \text{GeV}$   &  $v_{\nu_3}=1.2\times10^{-4} \text{GeV}$     \\ \hline 
       $|U_{\tilde{\gamma} \nu}|^2=2.16\times10^{-19}$ &  $Y_{\nu_{1}}=3\times10^{-7}$  &  $Y_{\nu_{2}}=0.9\times10^{-6}$  &   $Y_{\nu_{3}}=0.7\times10^{-6}$ \\ \hline
   \end{tabular}
    \caption{The same as in Table~\ref{param2} but for $M_2=M_1$.
}
\label{param1} 
\end{center}
\end{table}

\clearpage

\section{Conclusions}\label{sec:conclusions}

The $\mu\nu$SSM is a supersymmetric model that solves the $\mu$ problem and reproduces neutrino data, simply using couplings with the three families of right-handed neutrinos.
Since these couplings break $R$-parity, the gravitino is an interesting candidate for dark matter in this model.

In this work we have carried out a complete analysis of the detection of the gravitino as a decaying dark matter candidate in the $\mn$. In addition to the two-body decay (see Fig.~\ref{fig_gfn}) producing a sharp line with an energy at half of the gravitino mass,
we have included in the analysis the three-body decays (see Figs.~\ref{fig_gz} and~\ref{fig_gw}) producing a smooth spectral signature.
Then, we have compared the $\gamma$-ray fluxes predicted by the model with \Fermi LAT observations.
In particular, with the
95$\%$ CL upper limits on the total diffuse extragalactic $\gamma$-ray background (EGB) using 50 months of data, together with the upper limits on line emission from an updated analysis using
69.9 months of data.

We have performed first a deep exploration of the low-energy parameter space of the $\mn$ taking into account that neutrino data must be reproduced. This imposes important constraints, and, as a consequence, the photino-neutrino mixing parameter must fulfill the bounds 
$10^{-20} \lesssim|U_{\tilde{\gamma} \nu}|^2\lesssim 10^{-14}$. 
This parameter is crucial in the computation of the line, since 
the smaller $|U_{\tilde{\gamma} \nu_i}|^2$ 
the smaller is the decay width, and therefore less stringent are the constraints on the gravitino mass 
(see Eq.~\ref{decay2body})) from the non-observation of lines in \Fermi LAT data.
The relevant lower bound can be obtained for large values of the gaugino masses $M_{1,2}$ and/or in the limit
$M_2 \to M_1$ (see the approximate formula (\ref{UAP1})). 

We have found that in standard scenarios such as those with the low-energy GUT relation $M_2 = 2 \, M_1$, the line limits
are crucial and only allow gravitinos with masses $\lsim 4~\gev$ (and lifetimes $\gsim 10^{28}$ s), even for values of $|M_1|$ as large as 
10 TeV.
In the case $M_2 \to M_1$, although the line size is suppressed restricting less the 
gravitino mass ($\lsim 17$ GeV), still the line limits are more important than the EGB 
ones (see Fig.~\ref{fig_EGBa}). The latter only 
rule out the extreme scenarios, i.e. with very large and very close gaugino masses.
Nevertheless, the smooth signal from three-body decays plays an important role 
since can dominate the gravitino decay rate over a wide region of those parameters (see Fig.~\ref{fig2}), hence modifying the exclusion limits for its lifetime.

Our results are summarized in Figs.~\ref{figconstrains} and~\ref{figsymmetric12}, where we can see that using \Fermi LAT data the following constraints on the gravitino mass and lifetime in the $\mn$ are obtained:
$m_{3/2}\lsim 17~\gev$ and $\tau_{3/2}\gsim 4\times 10^{25}$ s.

\vspace{1cm}

\bigskip
{\bf Acknowledgments.} 
The work of GAGV 
was supported by Programa FONDECYT
Postdoctorado under grant 3160153. GAGV also thanks the IFT UAM-CSIC
for the hospitality during the completion of this work
.
The work of DL and AP was supported by the Argentinian CONICET.
The work of CM was supported in part by the Spanish grant FPA2015-65929-P MINECO/FEDER
UE, and by the Programme SEV-2012-0249 `Centro de Excelencia Severo Ochoa'. 
The work of R. RdA was supported by the Ram\'on y Cajal program of the Spanish MINECO and also thanks the support of the grants 
FPA2014-57816-P and FPA2013-44773, and the Severo Ochoa MINECO project SEV-2014-0398.
We also acknowledge specially the support of the Spanish MINECO's Consolider-Ingenio 2010 Programme under grant MultiDark CSD2009-00064.

\appendix

\section{Gravitino three-body decay channels}
\label{3body}
The calculation of the tree-level gravitino decays via an intermediate photon or $Z$ boson, and an
intermediate $W$ boson, was carried out in Refs.~\cite{Choi:2010xn,Choi:2010jt,Grefe:2011dp,Diaz:2011pc}. We show in this Appendix the results for the differential decay widths with respect to $s$~\cite{Grefe:2011dp}, where $s$ is the invariant mass of the two fermions, $f$ and $\bar{f}$.
The total decay widths can be obtained integrating these results
over the invariant mass range $0\leq s \leq m_{3/2}^2$. In the case of virtual photon exchange, one should integrate over the range $4m_f^2\leq s \leq m_{3/2}^2$, to avoid a divergent propagator.\\



%

\noindent
\textbf{i. $\Psi_{3/2}\rightarrow\gamma^*/Z^* \, \nu_i\rightarrow f \,  \bar{f} \, \nu_i$} 
\bea
\begin{aligned}
 &\frac{d\Gamma(\Psi_{3/2}\rightarrow \gamma^*/Z^* \, \nu_i\rightarrow f \,  \bar{f} \, \nu_i)}{ds}\\ 
 & \approx \frac{m_{3/2}^3 \, \beta_s^2}{768 \, \pi^3 \, M_{P}^2}\left[ \frac{e^2 \, Q^2}{s} \, |U_{\tilde{\gamma} \nu_i}|^2 \, f_s \, \sqrt{1-4 \, \frac{m_f^2}{s}}\left( 1+2 \, \frac{m_f^2}{s}\right) \right.\\
 & + \frac{g_Z}{(s-m_Z^2)^2+m_Z^2 \, \Gamma_Z^2} \left\lbrace \vphantom{\frac12} g_Z \,  U_{\tilde{Z} \nu_i}^2 \, s \, (C_V^2+C_A^2) \, f_s \right. \\
 & - \frac{8}{3} \, \frac{m_Z}{m_{3/2}} \, g_Z \,  U_{\tilde{Z} \nu_i}  \, \left( \frac{v_{\nu_i}}{v}+ \sin\beta \, \text{Re}  \, U_{\tilde{H}_u^0 \nu_i} \, - \cos\beta \, \text{Re} \,  U_{\tilde{H}_d^0 \nu_i} \, \right)  \,  s \, (C_V^2+C_A^2) \, j_s \\
 & + \frac{1}{6} \, g_Z  \, m_Z^2  \, \left|\frac{v_{\nu_i}}{v}+ \sin\beta \,  U_{\tilde{H}_u^0 \nu_i} \, - \cos\beta \,  U_{\tilde{H}_d^0 \nu_i} \, \right|^2 \, (C_V^2+C_A^2) \, h_s \\
 & + e \, Q \left(  \,  \text{Re} \,  U_{\tilde{\gamma} \nu_i} \, (m_Z^2-s) \, + \, \text{Im} \,  U_{\tilde{\gamma} \nu_i} \,  m_z  \, \Gamma_Z \right)  \,  C_V \\
 & \left. \left. \times \left( \,  2  \, U_{\tilde{Z} \nu_i} \, f_s \, +\frac{8}{3} \, \frac{m_Z}{m_{3/2}} \left( \,  \frac{v_{\nu_i}}{v}+ \sin\beta \, \text{Re}  \, U_{\tilde{H}_u^0 \nu_i} \, -  \, \cos\beta \, \text{Re} \,  U_{\tilde{H}_d^0 \nu_i} \,  \right)   \, j_s \, \right) \vphantom{\frac12} \right\rbrace  \vphantom{\sqrt{1-4\frac{m_f^2}{s}}} \right]\ ,
\label{width1}
\end{aligned}
\eea
where $g_Z=g_2 \, / \cos\theta_W$ is the gauge coupling of the $Z$ boson, $m_Z$ and $\Gamma_Z$ its mass and decay width into two fermions, $v$ is the Higgs VEV, $Q$ is the charge of the final state fermions, and $C_V$ and $C_A$ are the coefficients of the $V-A$ structure of the $Z$ boson vertex with two fermions
\bea
    C_V=\frac{1}{2}T^3-Q\sin^2\theta_W\ , \hspace{1cm} C_A=-\frac{1}{2}T^3\ . 
\label{}
\eea
The kinematic functions $\beta_s$, $f_s$, $j_s$ and $h_s$ are given by
\begin{equation*}
 \beta_s=1-\frac{s}{m_{3/2}^2}\ , \hspace{1cm} f_s=1+\frac{2}{3}\frac{s}{m_{3/2}^2} +\frac{1}{3}\frac{s^2}{m_{3/2}^4}\ ,
\end{equation*}

\bea
    j_s=1+\frac{1}{2}\frac{s}{m_{3/2}^2}\ , \hspace{1cm} h_s=1+10\frac{s}{m_{3/2}^2} +\frac{s^2}{m_{3/2}^4}\ . 
\label{kin}
\eea
Note that $U_{\tilde{\chi} \nu_i}$ denotes the mixing between the $\tilde{\chi}$ neutralino  and the $\nu_i$ neutrino, obtained from the neutral fermion (neutralino-neutrino) mass matrix.\\

\pagebreak

\noindent \textbf{ii. $\Psi_{3/2}\rightarrow W^* \, l\rightarrow f  \, \bar{f}' \, l$} 


\bea
\begin{aligned}
 &\frac{d\Gamma(\Psi_{3/2}\rightarrow W^{+*} \, l^-_i\rightarrow f \,  \bar{f}' \, l^-_i)}{ds}\\ 
 & \approx \frac{g_2^2  \, m_{3/2}^3 \,  \beta_s^2}{1536 \, \pi^3 \, M_{P}^2 \, \left( (s-m_W^2)^2+m_W^2 \, \Gamma_W^2\right) } \left(\vphantom{\frac12}  \, s \,  U^2_{\tilde{W}^- l_i^-}  \, f_s \right. \\
 & \left. -\frac{8}{3} \, \frac{m_W}{m_{3/2}} \, s \, U_{\tilde{W}^- l_i^-} \,  
\left( \frac{v_{\nu_i}}{v}-\sqrt{2} \,  \cos \beta  \,  \text{Re} \,  U_{\tilde{H}_d^- l_i^-} \, \right)  \,  j_s \,  + \frac{1}{6} \,  m_W^2  \, \left|\frac{v_{\nu_i}}{v}-\sqrt{2} \,  \cos \beta  \, U_{\tilde{H}_d^- l_i^-} \, \right|^2 \,  h_s  \, \vphantom{\frac12}\right)\ ,
\label{width2}
\end{aligned}
\eea
where $U_{\tilde{\chi}^- l_i^-}$ denotes the mixing between the $\tilde{\chi}^-$ chargino  and the $l_i^-$ lepton, obtained from the charged fermion (chargino-lepton) mass matrix. The kinematic functions $\beta_s$, $f_s$, $j_s$ and $h_s$ are the same as above.
\bibliography{glmpr-V2}
\bibliographystyle{JHEP}

\end{document}